\documentclass[%
 preprint, 
 amsmath,amssymb,
 aps, physrev,
floatfix,
]{revtex4-2}

\usepackage{graphicx}% Include figure files
\usepackage{dcolumn}% Align table columns on decimal point
\usepackage{bm}% bold math
\usepackage{multirow}
\begin{document}

\preprint{APS/123-QED}

\title{First-principles study of phase stability and magnetic properties of B2 AlCr, AlMn, AlFe, AlCo and AlNi aluminides}% Force line breaks with \\

\author{Haireguli Aihemaiti}
 \email{Contact author: haiaih@kth.se}
\affiliation{Department of Materials Science and Engineering, KTH Royal Institute of Technology, Stockholm SE-100 44, Sweden}% 
\affiliation{Wallenberg Initiative Materials Science for Sustainability, Department of Materials Science and Engineering, KTH Royal Institute of Technology}

\author{Esmat Dastanpour}%
\affiliation{Department of Materials Science and Engineering, KTH Royal Institute of Technology, Stockholm SE-100 44, Sweden}% 
\affiliation{Wallenberg Initiative Materials Science for Sustainability, Department of Materials Science and Engineering, KTH Royal Institute of Technology}

\author{Anders Bergman}%
\affiliation{Department of Physics and Astronomy, Uppsala University, Box 516, SE-75120 Uppsala, Sweden}% 

\author{Levente Vitos}%
\email{Contact author: leveute@kth.se}
\affiliation{Department of Materials Science and Engineering, KTH Royal Institute of Technology, Stockholm SE-100 44, Sweden}% 
\affiliation{Wallenberg Initiative Materials Science for Sustainability, Department of Materials Science and Engineering, KTH Royal Institute of Technology}
\affiliation{Department of Physics and Astronomy, Division of Materials Theory, Uppsala University, Uppsala SE-751 20, Sweden}

\begin{abstract}
Using $ab$ $initio$ Density Functional Theory (DFT) calculations, we investigate the electronic structure, phase stability, and magnetic properties of equiatomic binary alloys between Al and 3$d$ magnetic transition elements (Cr, Mn, Fe, Co, and Ni). Thermodynamically, all five binary aluminides are more stable in the ordered B2 phase than in the disordered body centered cubic phase, and Co is found to be the strongest B2 forming element with Al. The AlCo and AlNi compounds with B2 structure are verified to be non-magnetic, whereas AlFe turns out to be weakly magnetic, which is consistent with other DFT calculations employing similar exchange-correlation approximations. Magnetic simulations based on the Heisenberg Hamiltonian predict an antiferromagnetic ground state for the hypothetical B2 AlCr, which is also confirmed by direct DFT calculations. Doping AlCr with Co leads to an antiferromagnetic to ferromagnetic transition, where ferromagnetism is to a large extent attributed to Cr atoms. The phase stability and magnetic trends are explained using electronic structure arguments. The present findings contribute to a deeper understanding of the phase stability and magnetic properties of AlX binary alloys, providing insights into the formation mechanisms of the B2 structure with 3$d$ magnetic transition metals.
\begin{description}
\item[Keywords]
DFT, phase stability, magnetic properties, B2 structure, magnetic transition elements
\end{description}
\end{abstract}

\maketitle

%\tableofcontents

\section{\label{sec:level1}Introduction\protect\\}

The magnetic 3$d$ transition metals (Cr, Mn, Fe, Co, and Ni) have been given significant attention due to their intrinsic properties originating mainly from the $d$-electrons. The itinerant $d$-electrons are responsible for the specific phase stability, cohesive, and magnetic properties. Tailoring the magnetic state of these metals and their alloys by pressure, temperature or magnetic field is important for designing and developing new materials for various applications. 

The magnetic 3$d$ metals crystallize in structures that deviate from those of the 4$d$ and 5$d$ counterparts. At ambient conditions, Cr, Mn ($\alpha$-Mn), and Fe ($\alpha$-Fe) have body-centered cubic (bcc) structure, with 1 atom per primitive cell in the case of Cr and Fe, and 29 atoms per primitive cell in the case of Mn. The other two 3$d$ magnetic metals have close-packed structures: Co ($\beta$-Co) is hexagonal close-packed (hcp) and Ni is face-centered cubic (fcc).  Manganese is antiferromagnetic (AFM) below 95 K, while Fe, Co and Ni are ferromagnetic with Curie temperatures of 1043 K, 1400 K and 627 K, respectively \cite{PhysRevB.65.184432, kvashnin2016microscopic, cardias2017bethe}. At low temperatures, Cr has an incommensurate spin-density wave state with N\'eel temperature of 311 K. In theoretical modeling, the magnetic state of Cr is often approximated by an AFM state with B2 structure \cite{PhysRevB.65.184432}. The nearest-neighbor exchange interaction of bcc Cr is negative, while the second nearest-neighbor interaction is positive, leading to the AFM coupling in the B2 lattice \cite{cardias2017bethe, kvashnin2016microscopic}. 

Changing the chemical or structural environment of these metals can lead to changes in their magnetic state. A typical example is Fe, which shows ferromagnetic order in the bcc structure but becomes non-magnetic (NM) in the hcp structure \cite{zaoui2020competition}. Chromium at free surfaces exhibits a rather different magnetic behavior compared to bulk Cr \cite{punkkinen2011surface,kadas2009magnetism}. The local environment can also affect the high-temperature paramagnetic state, e.g. near the stacking faults in Fe-Cr-Ni alloys \cite{vitos2006evidence,li2016first}. In fact, manipulating the crystal structure and chemistry turned out to be a promising route to develop new rare-earth-free magnetic materials for specific magnetic applications, such as the magnetocaloric effect.

A previous theoretical investigation by Zelen\'y $et$ $al.$ \cite{PhysRevB.83.184424} on the magnetic properties of Fe, Co, and Ni along the trigonal deformation path revealed that these metals remain FM in fcc, simple cubic (sc) and bcc structures as well. However, to the best of our knowledge, except for polonium, no pure metals crystallize in the sc structure. Moreover, the bcc structure of Co and Ni has not been observed. Therefore, no experimental evidence could directly verify these interesting theoretical predictions.

The possibility of tailoring the magnetic state by incorporating non-magnetic components into these itinerant magnetic metals is another alternative to reach new magnetic materials with potential applications. A well-known case is FeRh with an ordered B2 structure formed by Fe and Rh sublattices. This compound exhibits a first-order AFM to FM metamagnetic transition around 340 K \cite{PhysRev.134.A1547, PhysRev.131.183, zarkevich2018ferh}. The transition is accompanied by a sizable magnetic moment change on the Rh sublattice and results in an exceptional magnetocaloric effect \cite{VIEIRA2021157811, PhysRevB.94.174435, SANCHEZVALDES2020166130}. Nevertheless, the criticality of Rh makes this system less promising for magnetocaloric applications.

In the present study, we concentrate on 3$d$ magnetic elements alloyed with 50 at.\% non-magnetic aluminum. Based on the equilibrium phase diagrams, we expect disordered bcc or ordered B2-type structures consisting of 50 at.\% magnetic metal in one sublattice and 50 at.\% Al in the other sublattice. Since Al is known as a strong bcc stabilizing element in steels and high-entropy alloys, it is assumed that by adding Al to the late 3$d$ transition metals, one can destabilize the hcp and fcc lattices relative to the bcc or B2 lattices. Indeed, according to the binary phase diagrams, stable B2 phases of AlFe, AlCo, and AlNi exist around the equiatomic compositions. In contrast, the B2 structure of AlCr and AlMn is not stable at equilibrium conditions. To understand why the B2 structure exists with the late 3$d$ metals but not with the early 3$d$ metals, we conduct a systematic study by focusing on the phase stability trend as we go from Cr to Ni. Investigating the hypothetical AlCr and AlMn with B2 structure provides valuable insight into the behavior of aluminides based on 3$d$ transition elements. The change in the properties of these binary alloys underscores the significant impact of crystal structure variations on the magnetic behavior of these elements.

Several recent publications \cite{DASTANPOUR2024173977, dastanpour2022structural, huang2023combinatorial} on Al-based high- and medium-entropy alloys showed possible candidates for environmentally friendly magnetocaloric materials. The main elements in those B2 structures are Al and Mn with a low fraction of Co. The strong magnetic response observed in AlMnCo-based alloys was primarily attributed to the presence of Mn, as both Al and Co are non-magnetic according to the theoretical calculations \cite{dastanpour2025structural}. Thus, to understand the mechanism driving the magnetic behavior, a thoughtful investigation of the underlying physics of the B2-type structures of these materials is essential. We adopt methods based on Density Functional Theory to reveal the phase stability and magnetic properties of equiatomic AlCr, AlMn, AlFe, AlCo and AlNi. First, we study the sc structures of the pure 3$d$ metals, which makes it possible to connect the present results to previous work based on the trigonal deformation path \cite{PhysRevB.83.184424}. Next, we consider the equatomic systems in base-centered cubic structures. The bcc structure represents the chemically disordered B2 phase, and it is stable at high temperatures in Al-Fe and Al-Mn systems. Here, we show that at static conditions, all five binary systems prefer the ordered B2 phase with respect to the bcc phase. We find a systematic trend in the magnetic behavior as we proceed from Cr to Ni. In particular, AlCo and AlNi are non-magnetic, while AlCr turns out to have an antiferromagnetic structure. Contrasting the results obtained for the sc lattice tells us about the effect of Al on the magnetic properties of the B2 compounds, in which the magnetic elements also form a sc sublattice. Finally, we explore the possibility of designing new ferromagnetic intermetallic compounds by combining the hypothetical antiferromagnetic AlCr with the non-magnetic AlCo. We show that in B2 Al(Cr$_{1-x}$Co$_x$) alloys with more than 22 at.\% Co, the ferromagnetic state is stable and magnetism is primarily due to Cr. 

The structure of the paper is as follows. In Section \ref{secII}, we describe the theoretical approach giving the most important numerical details. In Section \ref{secIII}, we present and discuss the results. 

\section{\label{secII}Theoretical approach\protect\\}

\subsection{Methodology}

Density Functional Theory (DFT) was used for the electronic structure and total energy calculations. The Kohn-Sham equations \cite{PhysRev.140.A1133} adopted the Local Density Approximation (LDA) \cite{PhysRevB.45.13244} and were solved using the Exact Muffin Tin Orbitals (EMTO) method \cite{PhysRevB.64.014107, vitos2007computational}. The alloy problem was described within the Coherent Potential Approximation (CPA) \cite{PhysRev.156.809, PhysRevB.5.2382, PhysRevLett.87.156401}. The total energy was obtained from the total non-spherical charge density according to the full charge density technique \cite{vitos1997full, kollar2000electronic, kissavos2006total}. The exchange-correlation part of the total energy was treated within the Perdew–Burke–Ernzerhof (PBE) gradient level approximation \cite{PhysRevLett.77.3865}. 

The $ab$ $initio$ calculations were performed for ferromagnetic (FM), non-magnetic (NM) and paramagnetic (PM) states. The PM state, in turn, was modeled using the disordered local magnetic moment (DLM) approach \cite{gyorffy1985first} in combination with CPA. For AlCr and Al(CrCo) alloys in B2 structure, we also carried out antiferromagnetic (AFM) calculations. The AFM structure was obtained from spin dynamic simulations adopting a Heisenberg Hamiltonian built from $ab$ $initio$ exchange interactions $J_{\text{ij}}$. These in turn we computed by the EMTO method using the generalized perturbation method as implemented in Ref. \cite{ruban2016atomic}.   

\subsection{Numerical details}

The bcc structure was treated as disordered, while in the B2 structure, Al occupied one sublattice and the 3$d$ metals or alloy the other sublattice. The equations of state (EOS) were established by computing the total energy of the alloy system for several volumes around the expected equilibrium volume. The equilibrium volume and bulk modulus were then determined by fitting the total energy results to a Morse-type function \cite{moruzzi1988calculated}. Within the EMTO method, the volume is described by the Wigner-Seitz radius $w$ defined as the radius of a sphere with the same volume as the volume per atom, $i.e.$, $\mathit{\Omega} =  (4\pi/3) w^3$ \cite{vitos2000application}. 20-40 different $w$ values were used to find a high quality EOS. The density of states (DOS), phase stability, magnetic state, and elastic constant calculations adopted the so obtained theoretical equilibrium volumes.  

For EOS calculations, a 13 × 13 × 13 uniform $k$-point mesh was set up in the irreducible part of the Brillouin zones of the sc, bcc and B2 lattices. The EMTO Green's functions were computed on a semicircular contour using 16 complex energy points. A 16-atom supercell was built to construct the AFM AlCr in the B2 structure. The B2 and bcc shear elastic constants $C'=(C_{11}-C_{12})/2$ and $C_{44}$ were obtained by orthorhombic and monoclinic deformations, respectively, adopting six distortions ($\delta = 0.00, 0.01, 0.02, 0.03, 0.04, 0.05$) \cite{vitos2007computational}. The $k$-mesh in the elastic constant calculations was increased to 39 × 39 × 39 for both orthorhombic and monoclinic unit cells.  For the AFM AlCr B2 structure, the $k$-points were set up as 11 × 11 × 11. The single-crystal elastic constants were extracted from the above shear elastic constants and the bulk modulus $B=(C_{11}+2C_{12})/3$.

To find the magnetic structure of AlCr, Monte Carlo simulations were carried out using the Hastings-Metropolis algorithm as implemented in the UppASD code \cite{Eriksson2017}. We considered system sizes of 16 × 16 × 16 unit cells for a temperature mesh of 10 K, where at each temperature, thermal equilibrium was obtained by performing 50 k Metropolis sweeps across the sample. The magnetic ground state was obtained by energy-minimization, $i.e.$, identifying the minimum energy $E(q)$ for trial helical spin-spirals with wavelength $q$, where $q$ was varied for a Monkhorst-Pack like 16 × 16 × 16 mesh within the first Brillouin zone.

\section{\label{secIII}Results and Discussion\protect\\}

\subsection{Equations of state for Cr, Mn, Fe, Co and Ni in sc structure}

\begin{figure*}[!htb]
    \centering
    \includegraphics[scale=0.11]{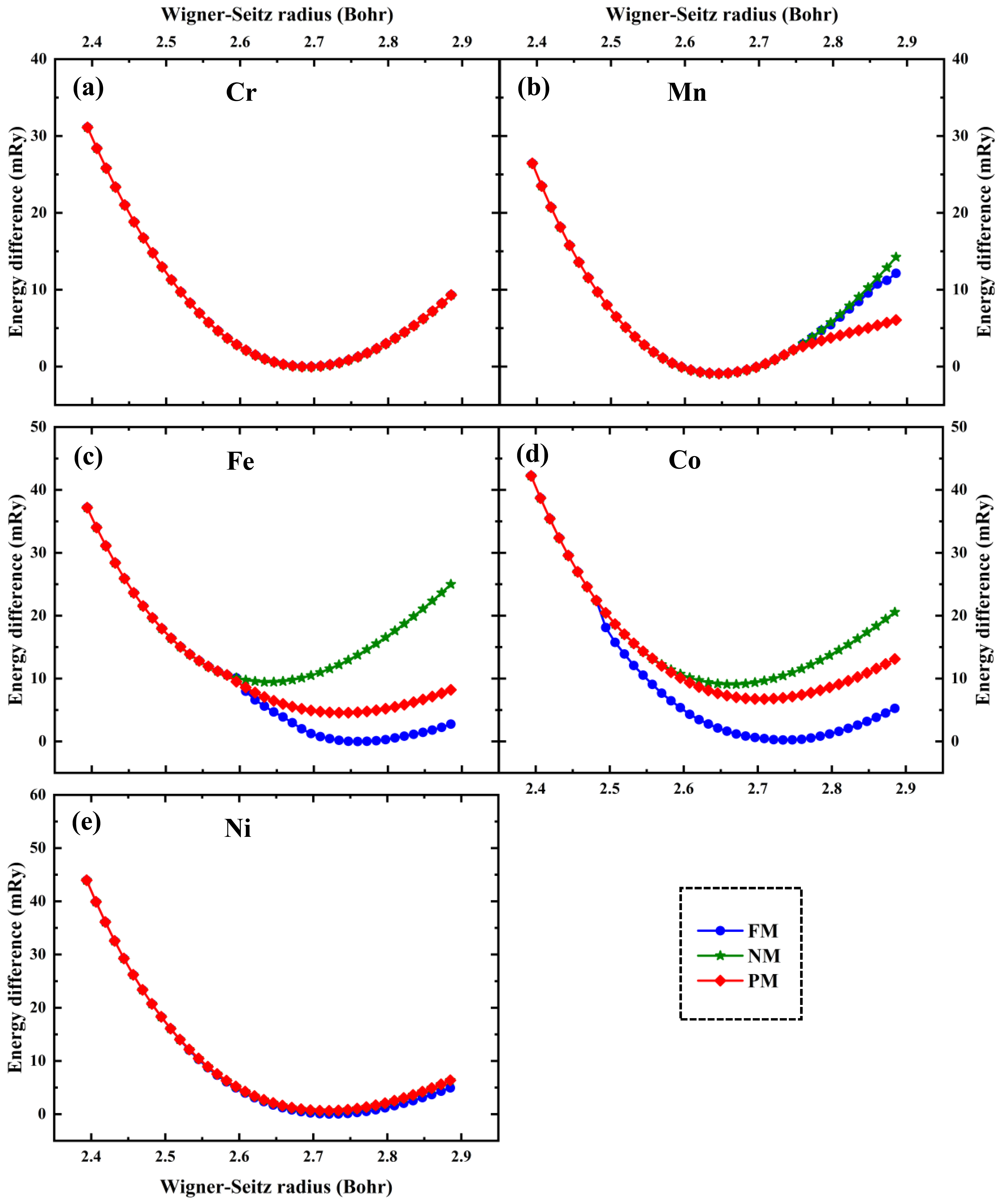}
    \caption{Theoretical total energies as a function of Wigner-Seitz radius for simple-cubic Cr (a), Mn (b), Fe (c), Co (d), and Ni (e). Results are shown for FM (blue), NM (green), and PM (red) states. In panel (a), the energy curves overlap for the three magnetic states.}
    \label{fig:1}
\end{figure*}
 
Figure \ref{fig:1} displays the EOS for the 3$d$ transition elements (Cr, Mn, Fe, Co, and Ni) in the sc structure. Results are shown for FM, NM, and PM states. It is observed that Cr has the same total energies versus Wigner-Seitz radius for all three magnetic states, meaning that sc Cr is NM. No antiferromagnetic structure was considered for pure sc Cr. Manganese in sc structure is also NM, but at high volumes, the PM energies are below the FM and NM energies, indicating that perhaps a non-collinear magnetic state develops at these volumes. Describing the high-volume magnetic state in sc Mn is beyond the scope of this study. 

The FM state is the most stable one for sc Fe, Co, and Ni, having the lowest energy near the corresponding equilibrium volumes. For these three elements, the PM state shows intermediate stability, while the NM state is the least stable one. These results are in line with those obtained by Zelen\'y $et$ $al.$ \cite{PhysRevB.83.184424}. They reported various magnetic states along the trigonal deformation path, but all three late transition metals were found to be in the FM state near the sc structure. The equilibrium volumes of sc Fe, Co, and Ni for the FM state are larger compared to the equilibrium volumes for the PM and NM states. This can be ascribed to the positive magnetic pressure present in itinerant magnets \cite{punkkinen2011compressive}. When contrasting the present FM volumes for sc lattice with those reported for bcc Fe, hcp Co and fcc Ni, we conclude that the sc structure yields larger equilibrium interatomic distances compared to the ground-state values \cite{vitos2007computational}. This finding is consistent with the one reported in Ref. \cite{PhysRevB.83.184424}. The good agreement with previous $ab$ $initio$ results verifies our theoretical approach based on the EMTO method and confirms that the present muffin-tin method correctly captures the magnetic states of the studied transition metals even for such open structures like the sc one.

\subsection{Equations of state for AlCr, AlMn, AlFe, AlCo and AlNi in bcc and B2 structures}

\begin{figure*}[!htb]
    \centering
    \includegraphics[scale=0.13]{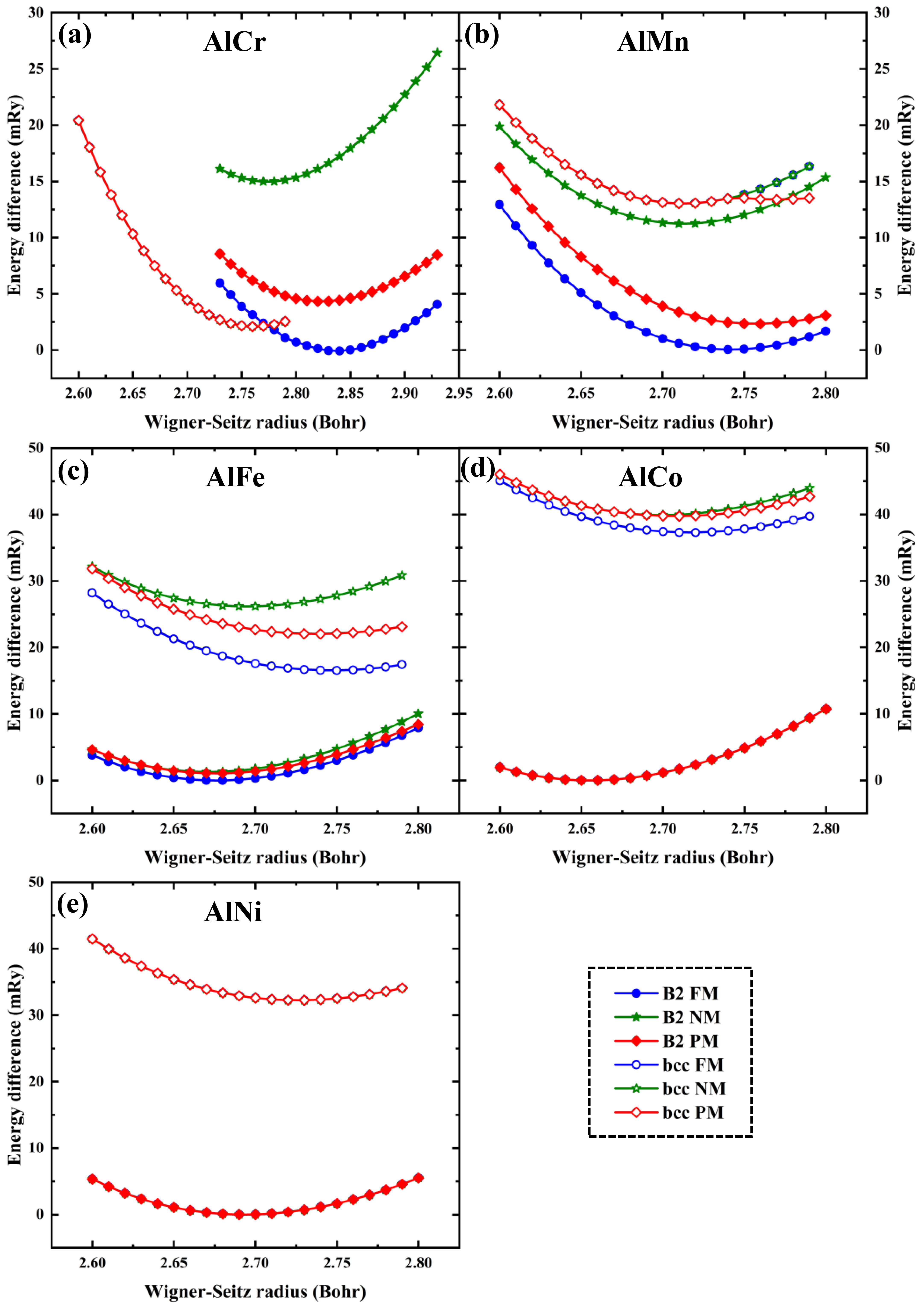}
    \caption{Theoretical total energies as a function of Wigner-Seitz radius for AlCr (a), AlMn (b), AlFe (c), AlCo (d) and AlNi (e) for  B2 (closed symbols) and bcc (open symbols) structures. Results are shown for FM (blue), NM (green), and PM (red) states. All curves are plotted relative to the B2 FM energy minimum. In panels (a) \& (e), the bcc energy curves overlap for the three magnetic states, while in panel (b), they overlap at Wigner-Seitz radii below 2.74 Bohr. In panels (d) \& (e), the B2 energy curves overlap with each other for all three magnetic states.}
    \label{fig:2}
\end{figure*}

Figure \ref{fig:2} shows the EOS for AlX (X= Cr, Mn, Fe, Co, and Ni) binary alloys in ordered B2 and disordered bcc structures. Results are given for FM, NM and PM states. The corresponding total energy differences, equilibrium Wigner-Seitz radii, total and local magnetic moments and magnetic transition temperatures are listed in Table \ref{table1}. Here, the Curie temperatures of the hypothetical FM phases were estimated using the mean-field approximation, according to which $T_{\text{C}}$ is proportional to the energy difference between the $E_\mathrm{PM}$ and $E_\mathrm{FM}$ states. Namely,  $T_{\text{C}}$ = $2(E_\mathrm{PM}$ - $E_\mathrm{FM})/3k_\mathrm{B} (1-c)$, where $c$ is the concentration of the non-magnetic component, and the energies are per atom. 

First, we analyze the stability of the bcc and B2 structures in various magnetic states and then compare the two sets of EOS. For the B2 structure, the EOS shows that the AlCr, AlMn, and AlFe alloys have the lowest energy in the FM state, indicating that for these alloys, the FM state is the most stable among the FM, PM and NM ones. The stability of the collinear magnetic configuration is especially pronounced for AlCr. That is, for AlCr, the FM energy minimum is below the NM energy minimum by $\sim$15 mRy per atom. The PM energies for AlCr and AlMn are also far below the NM energies, which means that there are sizable finite disordered local magnetic moments in these paramagnetic systems even at static conditions. Indeed, the local magnetic moment of Cr atoms in the FM and PM states is 2.59 $\mu_{\text{B}}$ and 2.29 $\mu_{\text{B}}$, respectively (Table \ref{table1}). We notice that the local magnetic moments on the Al sublattice are very small, and thus, we do not explicitly discuss them. For the AlCo and AlNi alloys, the FM, PM and NM energies coincide and the self-consistent local magnetic moments vanish for both the FM and PM systems, supporting the stability of the non-magnetic state. 

The total energy minima for the random bcc AlCr, AlMn, and AlNi alloys coincide for the three magnetic states, meaning that these systems are non-magnetic. For AlMn, the PM state becomes slightly more stable than the NM and FM states above the Wigner-Seitz radius of 2.74 Bohr. As for the sc structure, at those conditions, bcc AlMn could possess a non-collinear magnetic structure, which is not investigated here. The random bcc AlFe and AlCo alloys, on the other hand, are stable in the FM state with equilibrium local magnetic moments of 2.02 $\mu_{\text{B}}$ and 1.06 $\mu_{\text{B}}$ on the 3$d$ metal site, respectively. The PM EOS of these two systems possesses finite local magnetic moments (with equilibrium values of 1.67 $\mu_{\text{B}}$ and 0.44 $\mu_{\text{B}}$, respectively) and they are situated between the FM and NM curves.

By comparing the total energies of the B2 and bcc structures for AlCr, AlMn, AlFe, AlCo and AlNi, we find that the B2 phase consistently exhibits the lowest energy, and thus at static conditions it is the stable phase for all five systems with respect to the bcc phase. Looking at the structural energy differences ($\Delta E^{\rm FM}=E^{\rm FM}_\mathrm{bcc}-E^{\rm FM}_\mathrm{B2}$) between the bcc and B2 phases in the FM state, we find that AlCo has the largest energy difference (37.2 mRy), followed by AlNi (32.2 mRy), AlFe (16.5 mRy), AlMn (13.0 mRy), and AlCr (2.1 mRy). The large $\Delta E^{\rm FM}$ predicted for the AlCo alloy suggests that Co is the most efficient element among the 3$d$ elements considered here to stabilize the equiatomic alloy with Al in the ordered B2 structure with respect to the disordered bcc phase. These results are consistent with those found in previous $ab$ $initio$ calculations using the Linear Muffin-Tin Orbitals method \cite{kulikov1999onset}. It was reported that the formation energies of the ordered phases of AlFe, AlCo and AlNi are systematically lower than the formation energies of the corresponding disordered phases and that AlCo has 65\% and 55\% larger ordering energy than AlFe and AlNi, respectively. At this point, we should notice that CPA is a single-site approximation that neglects local lattice relaxations (LLRs). In the present case, the lattice mismatch is large between Al and 3$d$ metals, but it is small between the considered 3$d$ metals. Therefore, a sizable LLR error might be present in random bcc alloys (where Al is mixed with 3$d$ metals), but no LLR error is expected in ordered B2 structures. The energy difference between bcc and B2 will therefore be decreased by taking into account the LLR effects. However, since our results for AlNi, AlCo, AlFe and AlMn (doped with Co) are consistent with experiments, we have good reason to believe that the neglected LLR effects do not change the calculated phase stability trends \cite{vitos2007computational, huang2018mechanical}.

The equilibrium Al-Cr and Al-Mn binary phase diagrams reveal the absence of a stable B2 structure in equiatomic alloys. Interestingly, the Al-Cr system contains a very small region close to 60 at.\% Cr, where a B2-type of structure exists at high temperature \cite{helander1999experimental}. The samples used in those experiments were heat-treated at temperatures between 1160 and 1273 K and then water quenched. The measured room temperature lattice parameter of the B2-type compound was reported to be around 2.97 \AA, corresponding to a Wigner-Seitz radius of 2.76 Bohr. This experimental value is close to the present NM B2 or bcc Wigner-Seitz radii, but it is much smaller than our theoretical FM and PM radii (see Fig. \ref{fig:2}(a) and Table \ref{table1}). The main reason for the discrepancy is the chemical composition. The experimental B2 compounds have Cr content between 60 and 65 at.\% (read from Fig. 2 in Ref. \cite{helander1999experimental}) while the present calculations assume equiatomic composition. Using the experimental room-temperature Wigner-Seitz radii of Al (2.99 Bohr) and Cr (2.68 Bohr), Vegard's law predicts 2.79 Bohr for Al$_{0.35}$Cr$_{0.65}$ and 2.84 Bohr for Al$_{0.5}$Cr$_{0.5}$. The first value is close to the experimental one, while the latter one fits well the present equilibrium volume of magnetic AlCr.

Another reason for the discrepancy between the experimental and theoretical equilibrium volumes for the B2 AlCr compounds could be the magnetic state. Unfortunately, nothing was reported on the magnetic state of the observed B2 AlCr compound \cite{helander1999experimental}.  Figure \ref{fig:2}(a) displays that the FM B2 AlCr is much more stable than the NM or PM AlCr counterparts. A previous experimental-theoretical study \cite{dastanpour2025structural} confirms that the Al-Mn-X (X=Co, Fe, and Ni) ternary alloys have a B2 structure and are FM. The magnetic energy difference between the PM and FM states is 4.32 mRy for B2 AlCr, which should be compared to 2.29 mRy obtained for B2 AlMn. Thus, it is suggested that the B2 structure of AlCr is even more magnetic than the B2 AlMn alloy. This is reflected by the unusually large mean-field magnetic transition temperature (910 K) predicted for the hypothetical FM AlCr as compared to the one obtained for FM AlMn (483 K). Although such ferromagnetic behavior of B2 AlCr would be very appealing, one needs to carry out further magnetic modeling in order to find the true magnetic state of this system. This question will be further investigated in Section \ref{AFM-AlCr}.

Experimental phase diagrams for Al-Fe, Al-Co, and Al-Ni confirm that the stoichiometric B2 phases of these aluminides are non-magnetic. Our results verify the NM state for both AlCo and AlNi. The weak magnetization found for AlFe is in line with that reported in previous DFT modeling based on local density approximation \cite{kulikov1999onset}. The predicted FM ground state of B2 AlFe is an artifact of the local density or gradient level exchange-correlation approximations. Indeed, as discussed in Refs. \cite{mohn2001correlation,oestlin2017analytic}, using a non-local exchange-correlation schemes yields the non-magnetic ground state for AlFe within a wide volume range, in agreement with the experiment. Since the main focus here is on the phase stability and magnetic trends, we accept the small DFT error in the case of the AlFe system. 

Looking at Table \ref{table1}, we observe a clear trend in the total and local magnetic moments of the present alloys as we go from Cr to Ni. In the case of FM FeRh, a sizable magnetic moment is associated with the Rh sublattice effect \cite{VIEIRA2021157811, PhysRevB.94.174435, SANCHEZVALDES2020166130}. In the present FM B2 systems, the local magnetic moments are primarily connected to the 3$d$ metal sublattice and the Al sublattice remains nearly non-magnetic (Table \ref{table1}). The tiny local magnetic moments on Al atoms may in fact be due to the conventional Wigner-Seitz type of space distribution near the individual atoms. Namely, we assumed identical volumes associated with the 3$d$ metal site and Al site when integrating the magnetization density to obtain the local magnetic moments. The local magnetic moments of Cr (2.59 $\mu_{\text{B}}$ for FM and 2.29 $\mu_{\text{B}}$ for PM state) are larger than all other moments, even surpassing the value found for the Fe site in the bcc FM AlFe alloy (2.02 $\mu_{\text{B}}$). In the case of B2 AlMn, the FM and PM moments are around 1.90 and 2.09 $\mu_{\text{B}}$, respectively. As discussed above, the finite local moments in B2 AlFe are due to errors associated with the local density approximation \cite{oestlin2017analytic}. Both B2 AlCo and B2 AlNi exhibit zero local magnetic moments in both FM and PM states. In the random bcc lattice, only AlFe and AlCo have non-vanishing local magnetic moments, while bcc AlCr, AlMn and AlNi are non-magnetic. Finally, the estimated mean-field Curie temperatures of the FM phases are the largest for B2 AlCr (910 K) and bcc AlFe (1154 K), intermediate for B2 AlMn (483 K) and bcc AlCo (520 K) and small or zero for the rest. This trend follows the trend in the magnetic energy difference in Fig. \ref{fig:2}.
Interestingly, the measured and calculated Curie temperatures of the Co-stabilized B2 Al-Mn-Co alloys with Al content between 45 and 55 at.\% are around 391-511 K and 444-550 K, respectively \cite{dastanpour2025magnetocaloric}. The present result for stoichiometric B2 AlMn agrees well with these previously reported values. This finding confirms that magnetism in B2 Al(MnCo) is primarily due to Mn, while Co plays a phase stabilization effect.
 
\begin{table*}[!htb]
\caption{Theoretical Wigner-Seitz radius $w$ (Bohr), magnetic energy $E$ (mRy/atom), total magnetic moment $\mu_{total}$ ($\mu_{\text{B}}$/formula unit), local magnetic moment $\mu_{local}$ ($\mu_{\text{B}}$/atom), and Curie temperature $T_{\text{C}}$ (K) of AlCr, AlMn, AlFe, AlCo and AlNi alloys for the B2 and bcc structures in FM, NM, and PM state. All magnetic energies are shown relative to the minimum energy of the corresponding FM B2 phase.} \label{table1}
\centering
\def\arraystretch{0.5}%
\begin{ruledtabular}
\begin{tabular}{ccccccccc}

\textbf{Alloys}        & \textbf{Structure}   & \textbf{\begin{tabular}[c]{@{}c@{}}Magnetic \\ state\end{tabular}} & \textbf{$w$}          & \textbf{$E$}           & \textbf{$\mu_{total}$} & \multicolumn{2}{c}{\textbf{$\mu_{local}$}} & \textbf{$T_{\text{C}}$} \\ \hline
\multirow{10}{*}{AlCr} & \multirow{5}{*}{B2}  & \multirow{2}{*}{FM}                                                & \multirow{2}{*}{2.84} & \multirow{2}{*}{0}     & \multirow{2}{*}{2.65}  & \multicolumn{1}{c}{Al}       & 0.06        & \multirow{5}{*}{910}    \\ 
                       &                      &                                                                    &                       &                        &                        & \multicolumn{1}{c}{Cr}       & 2.59        &                         \\ 
                       &                      & NM                                                                 & 2.77                  & 14.99                 & 0                      & \multicolumn{2}{c}{0}                      &                         \\ 
                       &                      & \multirow{2}{*}{PM}                                                & \multirow{2}{*}{2.83} & \multirow{2}{*}{4.32} & \multirow{2}{*}{0}     & \multicolumn{1}{c}{Al}       & 0           &                         \\ 
                       &                      &                                                                    &                       &                        &                        & \multicolumn{1}{c}{Cr}       & 2.29        &                         \\ 
                       & \multirow{5}{*}{bcc} & \multirow{2}{*}{FM}                                                & \multirow{2}{*}{2.76} & \multirow{2}{*}{2.10}     & \multirow{2}{*}{0}     & \multicolumn{1}{c}{Al}       & 0           & \multirow{5}{*}{0}      \\ 
                       &                      &                                                                    &                       &                        &                        & \multicolumn{1}{c}{Cr}       & 0           &                         \\ 
                       &                      & NM                                                                 & 2.76                  & 2.10                      & 0                      & \multicolumn{2}{c}{0}                      &                         \\ 
                       &                      & \multirow{2}{*}{PM}                                                & \multirow{2}{*}{2.76} & \multirow{2}{*}{2.10}     & \multirow{2}{*}{0}     & \multicolumn{1}{c}{Al}       & 0           &                         \\ 
                       &                      &                                                                    &                       &                        &                        & \multicolumn{1}{c}{Cr}       & 0           &                         \\ 
\multirow{10}{*}{AlMn} & \multirow{5}{*}{B2}  & \multirow{2}{*}{FM}                                                & \multirow{2}{*}{2.74} & \multirow{2}{*}{0}     & \multirow{2}{*}{1.87}  & \multicolumn{1}{c}{Al}       & -0.03       & \multirow{5}{*}{483}    \\ 
                       &                      &                                                                    &                       &                        &                        & \multicolumn{1}{c}{Mn}       & 1.90        &                         \\ 
                       &                      & NM                                                                 & 2.71                  & 11.20                 & 0                      & \multicolumn{2}{c}{0}                      &                         \\ 
                       &                      & \multirow{2}{*}{PM}                                                & \multirow{2}{*}{2.76} & \multirow{2}{*}{2.29} & \multirow{2}{*}{0}     & \multicolumn{1}{c}{Al}       & 0           &                         \\ 
                       &                      &                                                                    &                       &                        &                        & \multicolumn{1}{c}{Mn}       & 2.09        &                         \\ 
                       & \multirow{5}{*}{bcc} & \multirow{2}{*}{FM}                                                & \multirow{2}{*}{2.71} & \multirow{2}{*}{13.00}     & \multirow{2}{*}{0}     & \multicolumn{1}{c}{Al}       & 0           & \multirow{5}{*}{20}     \\ 
                       &                      &                                                                    &                       &                        &                        & \multicolumn{1}{c}{Mn}       & 0           &                         \\ 
                       &                      & NM                                                                 & 2.71                  & 13.00                      & 0                      & \multicolumn{2}{c}{0}                      &                         \\ 
                       &                      & \multirow{2}{*}{PM}                                                & \multirow{2}{*}{2.71} & \multirow{2}{*}{13.00}     & \multirow{2}{*}{0}     & \multicolumn{1}{c}{Al}       & 0           &                         \\ 
                       &                      &                                                                    &                       &                        &                        & \multicolumn{1}{c}{Mn}       & 0.02        &                         \\ 
\multirow{10}{*}{AlFe} & \multirow{5}{*}{B2}  & \multirow{2}{*}{FM}                                                & \multirow{2}{*}{2.68} & \multirow{2}{*}{0}     & \multirow{2}{*}{0.71}  & \multicolumn{1}{c}{Al}       & -0.04       & \multirow{5}{*}{226}    \\ 
                       &                      &                                                                    &                       &                        &                        & \multicolumn{1}{c}{Fe}       & 0.75        &                         \\ 
                       &                      & NM                                                                 & 2.67                  & 1.27                  & 0                      & \multicolumn{2}{c}{0}                      &                         \\ 
                       &                      & \multirow{2}{*}{PM}                                                & \multirow{2}{*}{2.68} & \multirow{2}{*}{1.07} & \multirow{2}{*}{0}     & \multicolumn{1}{c}{Al}       & 0           &                         \\ 
                       &                      &                                                                    &                       &                        &                        & \multicolumn{1}{c}{Fe}       & 0.53        &                         \\  
                       & \multirow{5}{*}{bcc} & \multirow{2}{*}{FM}                                                & \multirow{2}{*}{2.75} & \multirow{2}{*}{16.51}     & \multirow{2}{*}{1.93}  & \multicolumn{1}{c}{Al}       & -0.09       & \multirow{5}{*}{1154}   \\ 
                       &                      &                                                                    &                       &                        &                        & \multicolumn{1}{c}{Fe}       & 2.02        &                         \\ 
                       &                      & NM                                                                 & 2.70                  & 26.14                   & 0                      & \multicolumn{2}{c}{0}                      &                         \\ 
                       &                      & \multirow{2}{*}{PM}                                                & \multirow{2}{*}{2.74} & \multirow{2}{*}{21.99}  & \multirow{2}{*}{0}     & \multicolumn{1}{c}{Al}       & 0           &                         \\ 
                       &                      &                                                                    &                       &                        &                        & \multicolumn{1}{c}{Fe}       & 1.67        &                         \\ 
\multirow{10}{*}{AlCo} & \multirow{5}{*}{B2}  & \multirow{2}{*}{FM}                                                & \multirow{2}{*}{2.66} & \multirow{2}{*}{0}     & \multirow{2}{*}{0}     & \multicolumn{1}{c}{Al}       & 0           & \multirow{5}{*}{0}      \\ 
                       &                      &                                                                    &                       &                        &                        & \multicolumn{1}{c}{Co}       & 0           &                         \\ 
                       &                      & NM                                                                 & 2.66                  & 0                      & 0                      & \multicolumn{2}{c}{0}                      &                         \\ 
                       &                      & \multirow{2}{*}{PM}                                                & \multirow{2}{*}{2.66} & \multirow{2}{*}{0}     & \multirow{2}{*}{0}     & \multicolumn{1}{c}{Al}       & 0           &                         \\ 
                       &                      &                                                                    &                       &                        &                        & \multicolumn{1}{c}{Co}       & 0           &                         \\ 
                       & \multirow{5}{*}{bcc} & \multirow{2}{*}{FM}                                                & \multirow{2}{*}{2.72} & \multirow{2}{*}{37.24}     & \multirow{2}{*}{1.01}  & \multicolumn{1}{c}{Al}       & -0.05       & \multirow{5}{*}{520}    \\ 
                       &                      &                                                                    &                       &                        &                        & \multicolumn{1}{c}{Co}       & 1.06        &                         \\ 
                       &                      & NM                                                                 & 2.70                  & 39.90                   & 0                      & \multicolumn{2}{c}{0}                      &                         \\ 
                       &                      & \multirow{2}{*}{PM}                                                & \multirow{2}{*}{2.71} & \multirow{2}{*}{39.71}  & \multirow{2}{*}{0}     & \multicolumn{1}{c}{Al}       & 0           &                         \\ 
                       &                      &                                                                    &                       &                        &                        & \multicolumn{1}{c}{Co}       & 0.44        &                         \\ 
\multirow{10}{*}{AlNi} & \multirow{5}{*}{B2}  & \multirow{2}{*}{FM}                                                & \multirow{2}{*}{2.69} & \multirow{2}{*}{0}     & \multirow{2}{*}{0}     & \multicolumn{1}{c}{Al}       & 0           & \multirow{5}{*}{0}      \\ 
                       &                      &                                                                    &                       &                        &                        & \multicolumn{1}{c}{Ni}       & 0           &                         \\ 
                       &                      & NM                                                                 & 2.69                  & 0                      & 0                      & \multicolumn{2}{c}{0}                      &                         \\ 
                       &                      & \multirow{2}{*}{PM}                                                & \multirow{2}{*}{2.69} & \multirow{2}{*}{0}     & \multirow{2}{*}{0}     & \multicolumn{1}{c}{Al}       & 0           &                         \\
                       &                      &                                                                    &                       &                        &                        & \multicolumn{1}{c}{Ni}       & 0           &                         \\ 
                       & \multirow{5}{*}{bcc} & \multirow{2}{*}{FM}                                                & \multirow{2}{*}{2.73} & \multirow{2}{*}{32.21}     & \multirow{2}{*}{0}     & \multicolumn{1}{c}{Al}       & 0           & \multirow{5}{*}{0}      \\ 
                       &                      &                                                                    &                       &                        &                        & \multicolumn{1}{c}{Ni}       & 0           &                         \\ 
                       &                      & NM                                                                 & 2.73                  & 32.21                      & 0                      & \multicolumn{2}{c}{0}                      &                         \\ 
                       &                      & \multirow{2}{*}{PM}                                                & \multirow{2}{*}{2.73} & \multirow{2}{*}{32.21}     & \multirow{2}{*}{0}     & \multicolumn{1}{c}{Al}       & 0           &                         \\ 
                       &                      &                                                                    &                       &                        &                        & \multicolumn{1}{c}{Ni}       & 0           &                         \\ 
\end{tabular}
\end{ruledtabular}
\end{table*}

\subsection{Stoner theory for bcc and B2 AlCr, AlMn, AlFe, AlCo and AlNi aluminides}\label{sec:DOSNM}

\begin{figure*}[!htb]
    \centering
    \includegraphics[scale=0.18]{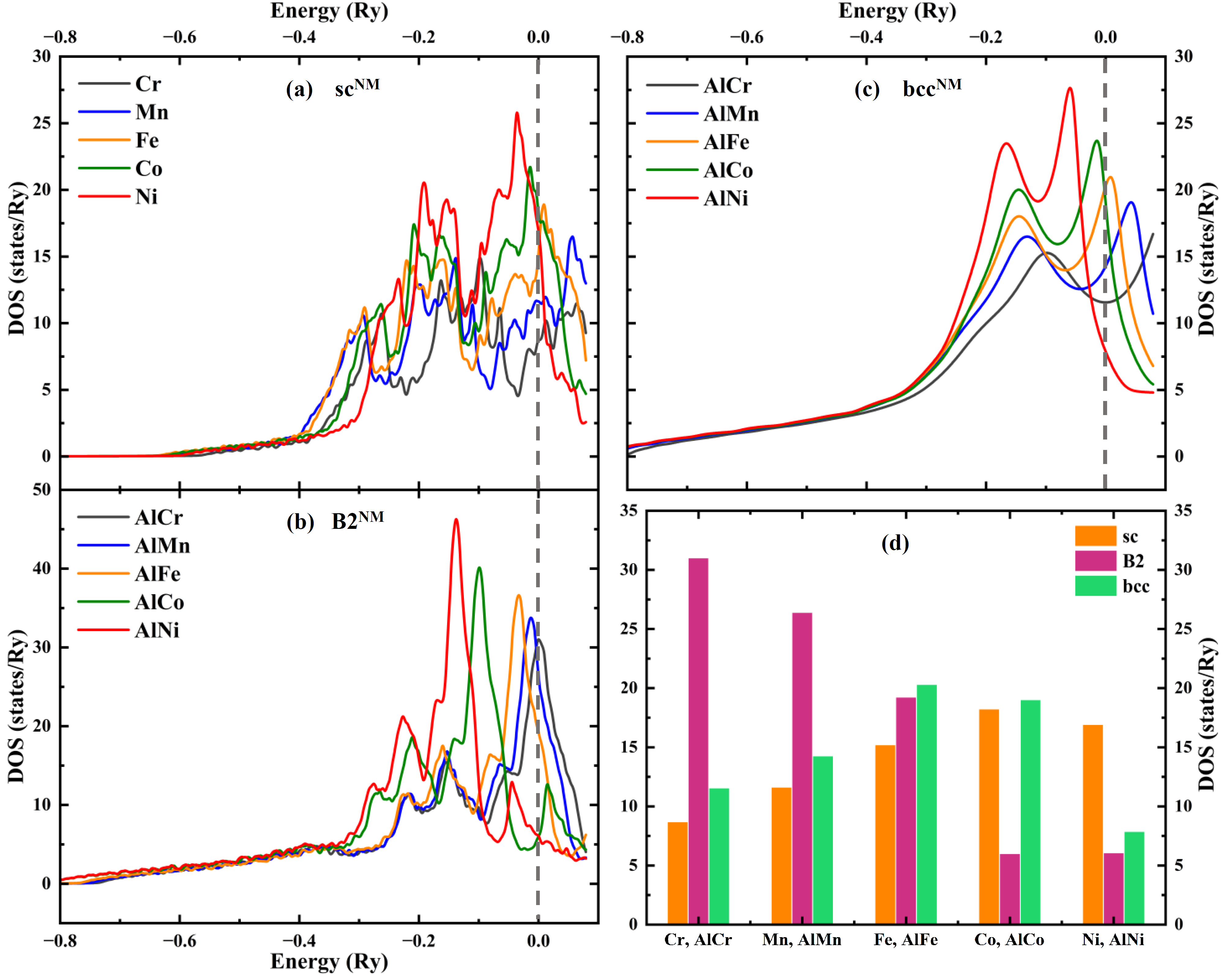}
    \caption{The calculated DOS of non-magnetic sc Cr, Mn, Fe, Co, and Ni for sc (a), B2 AlCr, AlMn, AlFe, AlCo, and AlNi (b), and bcc AlCr, AlMn, AlFe, AlCo, and AlNi (c). All calculations were carried out at the corresponding theoretical equilibrium volumes. Panel (d) collects the DOS values at the Fermi level for the three structures.}
    \label{fig:3}
\end{figure*}

To understand the magnetic phase stability of the present binary systems, we analyze the density of states (DOS) of the non-magnetic states. For completeness, in this analyses, we also include the sc structure. The DOS for NM sc Cr, Mn, Fe, Co, and Ni, and for NM B2 and NM bcc of AlCr, AlMn, AlFe, AlCo, and AlNi are presented in Fig. \ref{fig:3}, panels (a), (b) and (c). We observe that for all three structures, the DOS curves follow closely the rigid band model. Namely, the overall shape of the DOS curve is relatively unchanged for different elements, and only the position of the Fermi level moves up as we proceed from Cr to Ni. 

For the sc structure, as shown in Fig. \ref{fig:3}(a), the DOS values for Cr and Mn are relatively low at the Fermi level ($E_{\text{F}}$). According to the Stoner model \cite{stoner1938collective,stoner1939collective}, ferromagnetism occurs when the product of the exchange interaction parameter ($I$) and the NM DOS at the Fermi level, $N(E_{\text{F}})$, satisfies the Stoner criterion $IN(E_{\text{F}}) \geq 1$. For better accessibility, we collected the $N(E_{\text{F}})$ values for the three structures in Fig. \ref{fig:3}(d). To quantify the Stoner criterion, we take the exchange interaction parameters calculated from the Hartree-Fock (HF) method  \cite{oles1986influence} and local density approximation \cite{fritsche1998first}. For example, using $I$ = 0.08 Ry for Fe, the critical DOS at the Fermi level that leads to ferromagnetism is approximately 12.5 states/Ry per atom. Similarly, the HF $I$ values for Co and Ni \cite{oles1986influence} lead to 10.9 states/Ry and 10.3 states/Ry critical DOS values at the Fermi level, respectively. For Cr and Mn, the $I$ values obtained at the local density approximation level \cite{fritsche1998first} predict critical DOS values somewhat above that of Fe. Monitoring Fig. \ref{fig:3}(d), we find that the present sc DOS values for Cr and Mn are below and those for Fe, Co and Ni are above the corresponding critical values, and thus Stoner model predicts ferromagnetic state for sc Fe, Co and Ni but not for sc Cr and Mn. These findings are fully consistent with the total energy results presented in Fig. \ref{fig:1}. 

Figure \ref{fig:3}(b) displays the NM DOS curves for AlCr, AlMn, AlFe, AlCo, and AlNi alloys in the B2 structure. The DOS values at $E_{\text{F}}$ are significantly higher for AlCr, AlMn, and AlFe than those for AlCo and AlNi, confirming the ferromagnetic state of the first three ordered compounds. We notice that although AlFe follows a trend similar to that of AlCr and AlMn, its DOS value at $E_{\text{F}}$ is the smallest out of these three magnetic systems and relatively close to the critical value. The $N(E_{\text{F}})$ values for AlCo and AlNi are small, and thus, these two compounds are predicted to be NM. Interestingly, among the present B2 compounds, AlCo has the lowest NM DOS at $E_{\text{F}}$, suggesting that AlCo is the most stable B2 system, out of the present aluminides.

For the bcc structure, Fig. \ref{fig:3}(c), the DOS at $E_{\text{F}}$ is the highest for AlFe and AlCo, indicating tendencies for magnetic ordering. The DOS at $E_{\text{F}}$ for AlNi is significantly lower compared to those of AlFe and AlCo, suggesting that bcc AlNi remains NM. A comparative analysis of the B2 and bcc structures (Fig. \ref{fig:3}(d)) reveals that AlCr, which is NM in the bcc structure, becomes magnetic in the B2 structure. On the other hand, AlCo, which is magnetic in the bcc structure, turns NM in the B2 structure.

\subsection{Phase stability of FM AlCr, AlMn, AlFe, AlCo and AlNi aluminides}\label{sec:DOSFM}

The ternary Al-Mn-Co system was reported to have disordered bcc and ordered B2 phases \cite{dastanpour2025magnetocaloric}. The bcc phase fraction depends on Al and Co levels. In alloys with approximately 12 \% Co, the bcc volume faction varied between 0 and 41 \% as the Al content increased from 45 to 55 at.\%. In alloys with 50 at.\% Al, changing the Co level between 10 and 14 \% resulted in 9 to 23 \% bcc phase. Motivated by these earlier findings, in the following, we make an attempt to understand the competition between the bcc and B2 phases of the present aluminides. As discussed below, the stability of the B2 phase with respect to the bcc phase for the present aluminides has a complex electronic structure, electrostatic, and magnetic origin.

Phase stability can be established from the Gibbs energy at finite temperature or the total energy at static conditions. According to the force theorem \cite{mackintosh1980chapter,skriver1985crystal}, the total energy difference between similar structures can be estimated by comparing the one-electron energies. The one-electron energy, in turn, is the first-order momentum of the electronic DOS below the Fermi level. When the DOS is rearranged so that a local minimum develops near the Fermi level, the "missing" states are moved to lower energy. Therefore, phases with DOS minimum around the Fermi energy are preferred relative to the large DOS phases. Of course, the electrostatic contribution can modify the above simple picture \cite{skriver1985crystal}, but for the moment, we ignore that since we deal with similar structures.  

\begin{figure*}[!htb]
    \centering
    \includegraphics[scale=0.075]{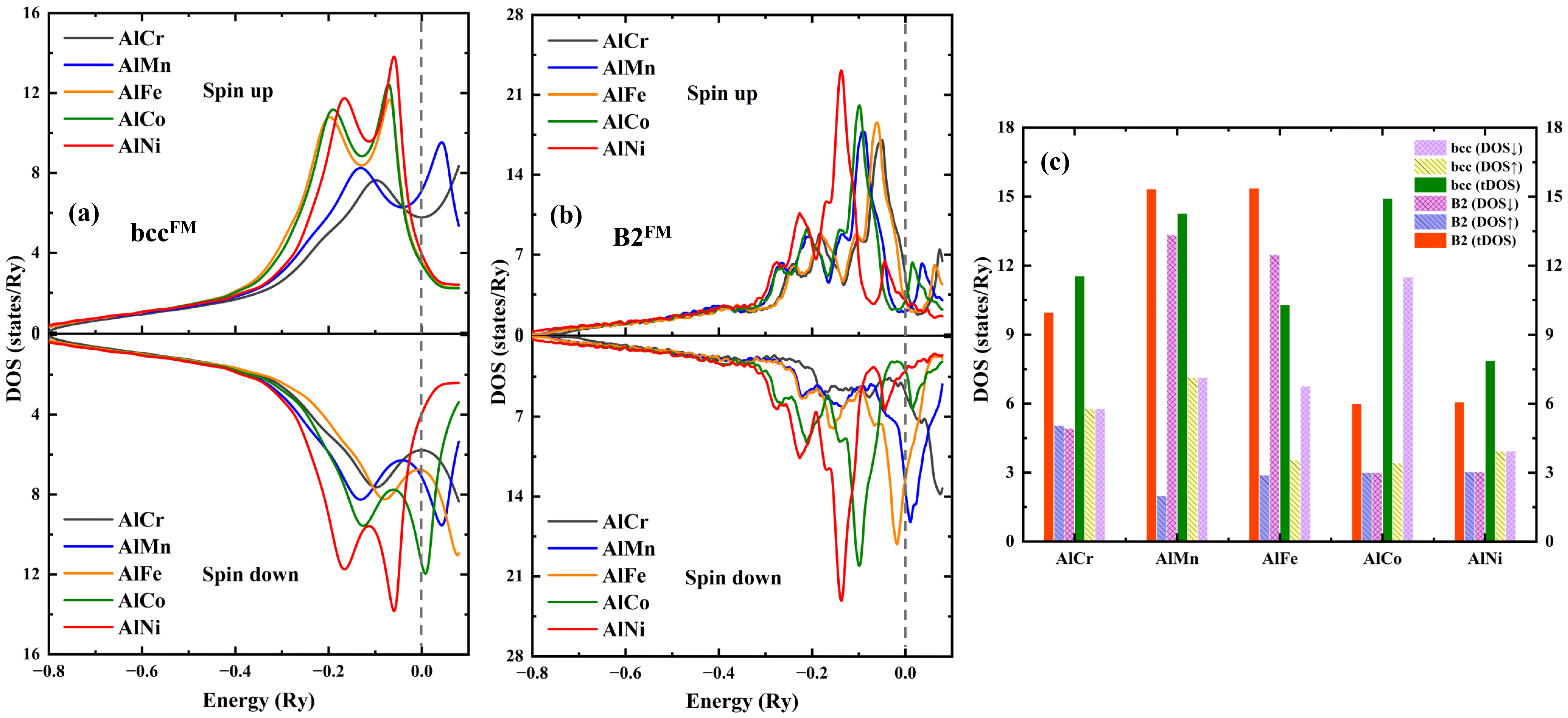}
    \caption{The calculated DOS of ferromagnetic  B2 AlCr, AlMn, AlFe, AlCo, and AlNi (a), and bcc AlCr, AlMn, AlFe, AlCo, and AlNi (b). All calculations were carried out at the corresponding theoretical equilibrium volumes. Panel (c) collects the DOS values at the Fermi level for the two structures.}
    \label{fig:4}
\end{figure*}

The FM DOS results for bcc and B2 AlCr, AlMn, AlFe, AlCo and AlNi are shown in Fig. \ref{fig:4} panels (a) and (b), respectively. In panel (c), we collect all total and spin FM DOS values at the Fermi level. For bcc AlCo, the DOS curve for spin-down electrons exhibits a prominent peak at the Fermi level, whereas for B2 AlCo, there is a local DOS minimum near the Fermi level in both spin channels. This suggests that for AlCo, the B2 phase is more stable compared to the bcc one. Similarly, although much smaller differences between the bcc and B2 DOS values at the Fermi level are observed also for AlCr and AlNi. These findings are in perfect line with Fig. \ref{fig:2} for AlCr, AlCo and AlNi. However, for AlMn and AlFe, the total DOS for the B2 phase is somewhat larger than that for the bcc one, although the differences are close to the numerical accuracy of our DOS calculation. Clearly, for these two systems, the thermodynamic stability of the B2 structures is not driven by the electronic states near the Fermi level but must be ascribed to the electrostatic and other effects. A simple estimation of the electrostatic contribution to the total energy is given by the Madelung expression \cite{skriver1985crystal}, according to which the negative energy correction is proportional to the Madelung constant and the square of the interstitial charge density. Accordingly, the larger the Madelung constant and the interstitial density are, the more stable the system is from electrostatic point of view \cite{korzhavyi1995madelung}. Since the Madelung constants for the present bcc and B2 structures are identical and the largest among all possible crystal lattices, one should look for the differences in the interstitial densities. For AlCr and AlMn, the bcc structure shows higher charge density in the interstitial regions, while for AlFe, AlCo, and AlNi, an opposite trend is observed (not shown). Therefore, from an electrostatic perspective, there is a clear trend: the B2 structure is favored for aluminades with late transition metals, while the bcc structure is preferred for aluminades with early transition metals. Therefore, the electrostatic contribution stabilizes the B2 phase against the bcc one for AlFe, AlCo and AlNi, which explains the pronounced phase stability seen in Fig. \ref{fig:2} for these three systems. Therefore, the particular features of the electronic structure near the Fermi level in combination with the electrostatic effect can explain the bcc-B2 phase stability for most of the aluminides considered here. AlMn seems to be an exception, where both of these effects predict bcc stability, in contrast to the DFT result in Fig. \ref{fig:2}. It is likely that for AlMn electronic states further below the Fermi level are responsible for the stability of the B2 phase, rather than the states near the Fermi level. Nevertheless, we must see that magnetism plays a crucial role in stabilizing the B2 phase over the bcc phase especially for AlMn and AlCr. As shown in Fig. 3(d), in the NM state, their DOS values at the Fermi level indicate that B2 is less stable than bcc. However, the magnetism-induced reduction of the B2 DOS near the Fermi level is particularly notable for these two systems. We end this discussion by concluding that based on the above electronic structure and electrostatic energy arguments, the AlCr, AlFe, AlCo and AlNi aluminides are thermodynamically more stable in the B2 structure than in the bcc structure, and the B2 AlCo system has the largest driving force for the phase stability. 

A natural next question is the dynamical stability of the predicted B2 phases of AlCr and AlMn. The elastic constants provide insights into the mechanical stability of particular crystal structures. We verify the dynamical stability of the present aluminides by computing the B2 and bcc elastic parameters at the corresponding equilibrium volumes. In a cubic lattice, there are three independent elastic constants: $C_{\text{11}}$, $C_{\text{12}}$, and $C_{\text{44}}$ \cite{vitos2007computational}. The criterion for the mechanical stability of a lattice requires that any small deformation results in a positive energy change. In terms of elastic parameters, a mechanically stable cubic system should meet the Born stability criteria \cite{born1940stability}: $C_{\text{11}} > 0$, $C_{\text{44}} > 0$, $C_{\text{11}} - C_{\text{12}} > 0$, and $C_{\text{11}} + 2C_{\text{12}} > 0$. The present results for FM bcc and B2 AlCr, AlMn, AlFe, AlCo and AlNi are summarized in Table \ref{table:2}.

Based on the present theoretical results, we conclude that all five aluminides are mechanically stable in the bcc and B2 structures. The Cauchy pressure ($C_{\text{12}} - C_{\text{44}}$) describes the angular nature of the atomic bonding. Namely, when $C_{\text{12}}$ is larger than $C_{\text{44}}$, the material tends to have metallic characteristics \cite{pettifor1992theoretical, zhang2018elastic}. The Cauchy pressures for the B2 structure are negative for AlCr, AlFe, and AlCo, indicating that in these systems, covalent bonding dominates over metallic behavior and they are intrinsically brittle. AlMn and AlNi show positive Cauchy pressures, supporting that they have good metallic behavior and are intrinsically ductile, especially AlMn having the largest positive Cauchy pressure \cite{temesi2024ductility}. In the B2 structure, AlFe has the highest $C_{\text{44}}$ value, while AlNi has the lowest one. Regarding the tetragonal shear modulus ($C' = ( C_{\text{11}} -  C_{\text{12}})/2$), AlMn shows the smallest value, indicating that this alloy is softer against tetragonal distortion than the other four alloys.

\begin{table}
\centering
\caption{Theoretical bulk moduli $B$ (GPa) and elastic constants $C_{\text{11}}$, $C_{\text{12}}$, $C_{\text{44}}$ (GPa) of AlCr, AlMn, AlFe, AlCo, and AlNi for the B2 and bcc structures in the FM state. Calculations correspond to the theoretical equilibrium Wigner-Seitz radii $w$ as shown in the third row (in Bohr). References for the previous theoretical results obtained using the PBE exchange-correlation approximation are indicated. } 
\label{table:2}

\setlength{\tabcolsep}{6pt}
\begin{tabular}{ccccccccccc}
\hline \hline 
\textbf{Alloys}                                                              
& \multicolumn{2}{c}{\textbf{AlCr}}                                                                   
& \multicolumn{2}{c}{\textbf{AlMn}}                                                                   
& \multicolumn{2}{c}{\textbf{AlFe}}                                                                              & \multicolumn{2}{c}{\textbf{AlCo}}                                                                              & \multicolumn{2}{c}{\textbf{AlNi}}   \\ \hline
Structure                                                                    & B2                                               & bcc                                              & B2                                               & bcc                                              & B2                                                                                        & bcc                                              & B2                                                                                        & bcc                                              & B2                                                                                             & bcc                                              \\
$w$                                                                          & 2.84                                             & 2.76                                             & 2.74                                             & 2.71                                             & 2.68                                                                                      & 2.75                                             & 2.66                                                                                      & 2.72                                             & 2.69                                                                                           & 2.72                                             \\
\begin{tabular}[c]{@{}c@{}}$B$ \\ Other theory (PBE)\end{tabular}            & \begin{tabular}[c]{@{}c@{}}138\\ -\end{tabular}  & \begin{tabular}[c]{@{}c@{}}172\\ -\end{tabular}  & \begin{tabular}[c]{@{}c@{}}156\\ -\end{tabular}  & \begin{tabular}[c]{@{}c@{}}176\\ -\end{tabular}  & \begin{tabular}[c]{@{}c@{}}172\\ 184\cite{chen2016elastic}\end{tabular}  & \begin{tabular}[c]{@{}c@{}}134 \\ -\end{tabular} & \begin{tabular}[c]{@{}c@{}}172\\ 181\cite{pagare2015first}\end{tabular}  & \begin{tabular}[c]{@{}c@{}}145\\ -\end{tabular}  & \begin{tabular}[c]{@{}c@{}}158\\ 159\cite{ponomareva2014effect}\end{tabular}  & \begin{tabular}[c]{@{}c@{}}144\\ -\end{tabular}  \\
\begin{tabular}[c]{@{}c@{}}$C_{\text{11}}$\\ Other theory (PBE)\end{tabular} & \begin{tabular}[c]{@{}c@{}}234 \\ -\end{tabular} & \begin{tabular}[c]{@{}c@{}}245 \\ -\end{tabular} & \begin{tabular}[c]{@{}c@{}}187 \\ -\end{tabular} & \begin{tabular}[c]{@{}c@{}}248 \\ -\end{tabular} & \begin{tabular}[c]{@{}c@{}}285 \\ 274\cite{chen2016elastic}\end{tabular} & \begin{tabular}[c]{@{}c@{}}170 \\ -\end{tabular} & \begin{tabular}[c]{@{}c@{}}312 \\ 286\cite{pagare2015first}\end{tabular} & \begin{tabular}[c]{@{}c@{}}173 \\ -\end{tabular} & \begin{tabular}[c]{@{}c@{}}233 \\ 233\cite{ponomareva2014effect}\end{tabular} & \begin{tabular}[c]{@{}c@{}}155 \\ -\end{tabular} \\
\begin{tabular}[c]{@{}c@{}}$C_{\text{12}}$\\ Other theory (PBE)\end{tabular} & \begin{tabular}[c]{@{}c@{}}89\\ -\end{tabular}   & \begin{tabular}[c]{@{}c@{}}136\\ -\end{tabular}  & \begin{tabular}[c]{@{}c@{}}141\\ -\end{tabular}  & \begin{tabular}[c]{@{}c@{}}141 \\ -\end{tabular} & \begin{tabular}[c]{@{}c@{}}115 \\ 139\cite{chen2016elastic}\end{tabular} & \begin{tabular}[c]{@{}c@{}}116 \\ -\end{tabular} & \begin{tabular}[c]{@{}c@{}}102 \\ 130\cite{pagare2015first}\end{tabular} & \begin{tabular}[c]{@{}c@{}}131 \\ -\end{tabular} & \begin{tabular}[c]{@{}c@{}}120 \\ 121\cite{ponomareva2014effect}\end{tabular} & \begin{tabular}[c]{@{}c@{}}139\\ -\end{tabular}  \\
\begin{tabular}[c]{@{}c@{}}$C_{\text{44}}$\\ Other theory (PBE)\end{tabular} & \begin{tabular}[c]{@{}c@{}}128\\ -\end{tabular}  & \begin{tabular}[c]{@{}c@{}}132\\ -\end{tabular}  & \begin{tabular}[c]{@{}c@{}}118\\ -\end{tabular}  & \begin{tabular}[c]{@{}c@{}}143 \\ -\end{tabular} & \begin{tabular}[c]{@{}c@{}}145 \\ 151\cite{chen2016elastic}\end{tabular} & \begin{tabular}[c]{@{}c@{}}120 \\ -\end{tabular} & \begin{tabular}[c]{@{}c@{}}138 \\ 253\cite{pagare2015first}\end{tabular} & \begin{tabular}[c]{@{}c@{}}123 \\ -\end{tabular} & \begin{tabular}[c]{@{}c@{}}116 \\ 114\cite{ponomareva2014effect}\end{tabular} & \begin{tabular}[c]{@{}c@{}}113 \\ -\end{tabular} \\
\hline \hline
\end{tabular}

\end{table}
\raggedbottom

\subsection{Magnetism of the B2 AlCr alloy}\label{AFM-AlCr}

Using $ab$ $initio$ calculations, we predict a magnetic B2 phase for the stoichiometric AlCr, which is not present in the equilibrium phase diagrams. The magnetic properties obtained so far are very appealing. Namely, the hypothetical FM AlCr system has large local magnetic moments associated with the Cr sites and a critical temperature of around 910 K. Such properties would make this metastable alloy very attractive for several potential applications. However, pure Cr is known to have an antiferromagnetic structure, which is driven by the strong antiferromagnetic nearest-neighbor exchange interactions \cite{cardias2017bethe, kvashnin2016microscopic}. This makes us look further into the magnetic structure of the B2 AlCr alloy.

\begin{figure*}[!htb]
    \centering
    \includegraphics[scale=0.09]{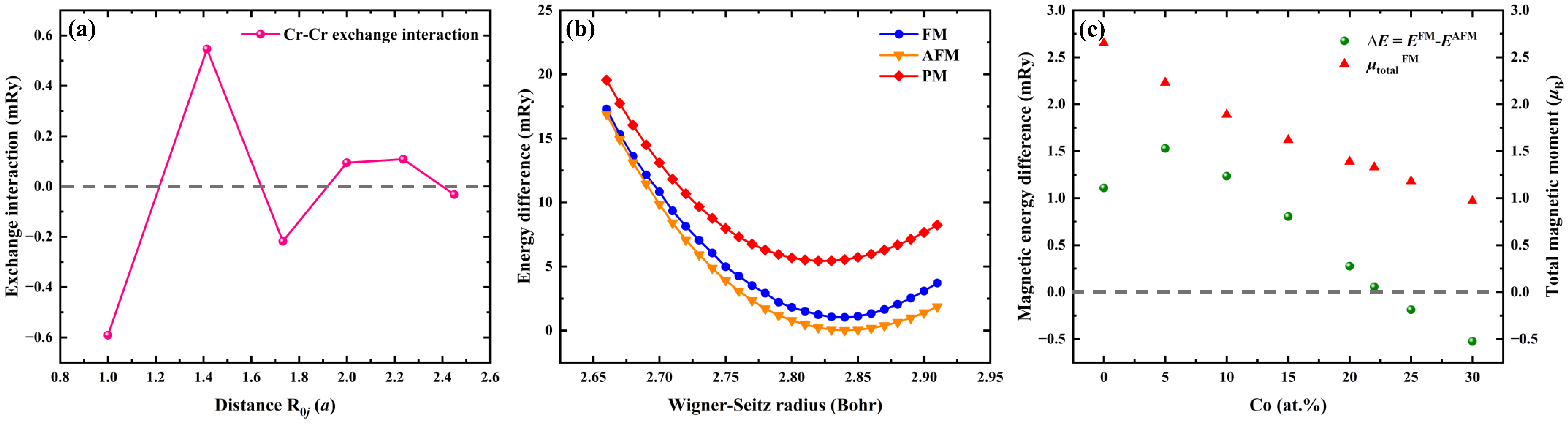}
    \caption{(a) The calculated Cr-Cr exchange interaction $J_{\text{0j}}$ for B2 AlCr. $\mathrm{R}_{\mathrm{0}j}$ represents the distance between atom $j$ and atom at the origin in units of B2 lattice parameter $a$. (b) The theoretical total energies of AlCr for the B2 structure in FM, AFM, and PM states. The FM and PM curves are identical to those in Fig. \ref{fig:2} and all energies are plotted relative to the AFM minimum. (c) The calculated magnetic energy difference ($\Delta E$ = $E^{\mathrm{FM}}$-$E^{\mathrm{AFM}}$) and total magnetic moment (per formula unit) in the FM state for B2 $\text{Al}_{50}\text{Cr}_{1-x}\text{Co}_x$ as a function of Co content.}
    \label{fig:5}
\end{figure*}

Starting from the FM state, we calculated the magnetic exchange interactions, $J_{\text{ij}}$, for AlCr. Since the Al sublattice is almost non-magnetic, the Al-Al and Al-Cr exchange interactions are very small, and thus we do not show them. The Cr-Cr magnetic exchange interactions for the sc Cr sublattice are shown in Fig. \ref{fig:5}(a) as a function of distance. We show only the first six interactions since the rest are negligible. It is found that the first-neighbor interaction is negative, while the second-neighbor interaction is positive, which is similar to the behavior observed for pure bcc Cr. This alternating sign of the first- and second-neighbor interactions suggests an AFM configuration for the B2 AlCr alloy.

\begin{figure*}[!htb]
    \centering
    \includegraphics[scale=0.09]{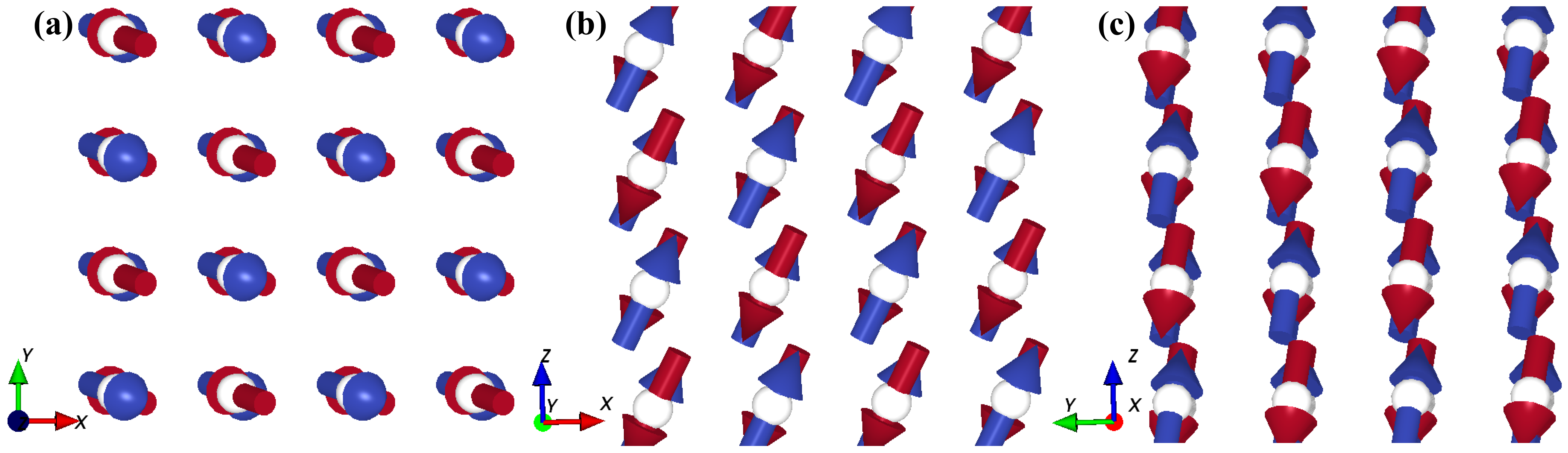}
    \caption{Illustration of the collinear antiferromagnetic order found in B2 AlCr, shown in the (a) $x$-$y$ plane, (b) $x$-$z$ plane, and (c) $y$-$z$ plane. Only the moments for the Cr atoms are shown. The magnetic structure is defined by an ordering vector $q$ = $(0.5, 0.5, 0.5)$. Since no magnetic anisotropy is taken into account, there is no preferred quantization axis in the system, thus, the spins appear tilted. The color scheme is chosen to maximize the contrast between the two existing spin directions.}
    \label{fig:6}
\end{figure*}

Using the so calculated magnetic exchange interactions, we constructed a Heisenberg Hamiltonian and carried out a Monte Carlo simulation from zero to 1000 K. We investigated the static structure factor $S(q)$, which can indicate if there are one or more ordering vectors in the system. We find a clear peak for $q=(0.5, 0.5, 0.5)$ which gives a collinear AFM ordering, where the spins flip direction every other atom in $x$, $y$, and $z$-directions. We visualize the predicted magnetic ordering at 1 K in Fig. \ref{fig:6}.

To confirm the AFM state in B2 AlCr predicted by the above magnetic simulations, we perform EOS calculations and compare the results with those obtained for the FM and PM states. As shown in Fig. \ref{fig:5}(b), the AFM state exhibits the lowest energy at the equilibrium volume, verifying its marginal stability with respect to the FM state. We also observe a small difference between the curvatures of the FM and AFM EOS, meaning that the present AFM state is slightly softer compared to the FM one. Interestingly, the theoretical equilibrium Wigner-Seitz radius (2.84 Bohr) of the AFM state is similar to the one obtained for the FM state (2.84 Bohr) and both are close to the value estimated from Vegard's law (2.84 Bohr). We also investigated the dynamical stability of the B2 AFM AlCr and calculated the following elastic constants: $B$ = 117 GPa,  $C_{\text{11}}$ = 165 GPa,  $C_{\text{12}}$ = 93 GPa and  $C_{\text{44}}$ = 114 GPa.  Our results confirm that the predicted B2 AFM AlCr is dynamically stable.

\subsection{Ferromagnetic B2 $\text{Al}_{50}\text{Cr}_{1-x}\text{Co}_x$ alloys}

The B2 phase of the ternary Al-Mn-Co system was stabilized by Co \cite{dastanpour2025magnetocaloric}. Actually, Co was found to partition into the B2 phase, and the random bcc phase had a reduced Co level compared to the average Co content. The total energy difference between bcc and B2 phases is larger for AlMn (13.0 mRy) than for AlCr (2.1 mRy). Hence, one might expect that the marginal stability of the B2 AlCr at static conditions can be enhanced by Co addition. Another aspect is magnetism. We do not know the low temperature magnetic state of AlMn. However, when Co was added to AlMn, it stabilized the FM state of the B2 phase. Therefore, by stabilizing the ordered B2 AlCr phase, one might also stabilize the FM state against the AFM one. In the following, we follow the above alloying strategy and investigate the magnetic energy difference in Co-doped AlCr as a function of composition. 

For this study, we add Co to the Cr sublattice of the B2 structure to explore how Co addition affects phase stability. The assumption that Co partitions to the Cr sublattice is motivated by previous calculations and measurements. Huang $et$ $al.$ \cite{huang2018strengthening} reported that in the partially ordered B2 structure of Al-doped Cr-Fe-Co-Ni high-entropy alloys, Al and Co occupy different sublattices. The theoretical-experimental work on Al-Mn-Co also supports this finding \cite{dastanpour2025magnetocaloric}. 

Using the EMTO method, we calculated the total energy of FM and AFM $\text{Al}_{50}\text{Cr}_{1-x}\text{Co}_x$ alloys as a function of concentration for $x\leq 30$ at.\%. For each concentration, the total energy is defined as the energy at the corresponding theoretical equilibrium volume. For the AFM state, we used the one described in the previous section. In Fig. \ref{fig:6}(c), we plot the difference between the FM and AFM total energies per atom. We find that when 5 at.\% Co is added to AlCr, the magnetic energy increases from about 1 mRy per atom to 1.5 mRy per atom. More Co addition decreases the magnetic energy difference and above approximately 22 at.\% Co, the B2 $\text{Al}_{50}\text{Cr}_{1-x}\text{Co}_x$ alloy stabilizes in the FM state with respect to the present AFM configuration. Interestingly, the total magnetic moment of the Co-doped FM system decreases with Co addition but remains above $\sim 1.4$ $\mu_{\text{B}}$ per Cr atom around the AFM-FM transition. Most importantly, in the FM $\text{Al}_{50}\text{Cr}_{1-x}\text{Co}_x$ alloys (with $x\gtrsim 22$ at.\%), the Co sites have negligible moments and thus the magnetization is almost entirely due to the Cr atoms. In other words, the otherwise antiferromagnetic Cr realizes a ferromagnetically ordered metal in the B2 Al-Cr-Co ternary alloy.

\section{Conclusion\protect\\}

We used Density Functional Theory formulated within the Exact Muffin-Tin Orbitals method in combination with the Coherent Potential Approximation to study a series of known and unknown aluminides formed by Cr, Mn, Fe, Co and Ni. For reference, we also carried out a systematic study of these 3$d$ metals in the hypothetical sc lattices. For aluminides, random bcc and ordered B2 phases were considered. 

The sc Fe, Co and Ni follow the previously reported trends. The agreement between observation and present predictions for AlCo and AlNi is also good. For AlFe, we incorrectly predict the ferromagnetic ground state, which is in line with previous results and can be ascribed to the local exchange-correlation approximation. Chromium is found to be nonmagnetic in the sc structure, but if it is forced to stay in the B2 structure, the hypothetical FM AlCr shows a large magnetization which is due to the pronounced peak in the non-magnetic density of state around the Fermi level. Further magnetic simulations demonstrate, however, an antiferromagnetic ground state for B2 AlCr. Since Co is the strongest B2 stabilizing element, we predict that the Co-doped AlCr gradually undergoes a magnetic transition from an antiferromagnetic state to a ferromagnetic state. The magnetism in ferromagnetic Al-Cr-Co is due to Cr atoms and reaches a theoretical value as high as $\sim 1.4$ $\mu_{\text{B}}$ per Cr atom. We call for experimental works that can synthesize this magnetic intermetallic compound. The predicted ferromagnetic B2 Al(CrCo) system is a promising candidate for magnetocaloric materials resembling the well-known B2 FeRh prototype. 

\section{Acknowledgments\protect\\}

This work was financially supported by the Wallenberg Initiative Materials Science for Sustainability (WISE) funded by the Knut and Alice Wallenberg Foundation (KAW). We also acknowledge the Swedish Foundation for Strategic Research, the Carl Tryggers Foundation, eSSENCE and Formas – a Swedish Research Council for Sustainable Development. The computations were enabled by resources provided by the National Academic Infrastructure for Supercomputing in Sweden (NAISS) at the National Supercomputing Centre (NSC, Tetralith cluster), partially funded by the Swedish Research Council through Grant Agreement No. 2022-06725.

\nocite{*}

\bibliography{apssamp}% Produces the bibliography via BibTeX.

\begin{thebibliography}{57}%
\makeatletter
\providecommand \@ifxundefined [1]{%
 \@ifx{#1\undefined}
}%
\providecommand \@ifnum [1]{%
 \ifnum #1\expandafter \@firstoftwo
 \else \expandafter \@secondoftwo
 \fi
}%
\providecommand \@ifx [1]{%
 \ifx #1\expandafter \@firstoftwo
 \else \expandafter \@secondoftwo
 \fi
}%
\providecommand \natexlab [1]{#1}%
\providecommand \enquote  [1]{``#1''}%
\providecommand \bibnamefont  [1]{#1}%
\providecommand \bibfnamefont [1]{#1}%
\providecommand \citenamefont [1]{#1}%
\providecommand \href@noop [0]{\@secondoftwo}%
\providecommand \href [0]{\begingroup \@sanitize@url \@href}%
\providecommand \@href[1]{\@@startlink{#1}\@@href}%
\providecommand \@@href[1]{\endgroup#1\@@endlink}%
\providecommand \@sanitize@url [0]{\catcode `\\12\catcode `\$12\catcode `\&12\catcode `\#12\catcode `\^12\catcode `\_12\catcode `\%12\relax}%
\providecommand \@@startlink[1]{}%
\providecommand \@@endlink[0]{}%
\providecommand \url  [0]{\begingroup\@sanitize@url \@url }%
\providecommand \@url [1]{\endgroup\@href {#1}{\urlprefix }}%
\providecommand \urlprefix  [0]{URL }%
\providecommand \Eprint [0]{\href }%
\providecommand \doibase [0]{https://doi.org/}%
\providecommand \selectlanguage [0]{\@gobble}%
\providecommand \bibinfo  [0]{\@secondoftwo}%
\providecommand \bibfield  [0]{\@secondoftwo}%
\providecommand \translation [1]{[#1]}%
\providecommand \BibitemOpen [0]{}%
\providecommand \bibitemStop [0]{}%
\providecommand \bibitemNoStop [0]{.\EOS\space}%
\providecommand \EOS [0]{\spacefactor3000\relax}%
\providecommand \BibitemShut  [1]{\csname bibitem#1\endcsname}%
\let\auto@bib@innerbib\@empty
%</preamble>
\bibitem [{\citenamefont {Hafner}\ \emph {et~al.}(2002)\citenamefont {Hafner}, \citenamefont {Spi\ifmmode~\check{s}\else \v{s}\fi{}\'ak}, \citenamefont {Lorenz},\ and\ \citenamefont {Hafner}}]{PhysRevB.65.184432}%
  \BibitemOpen
  \bibfield  {author} {\bibinfo {author} {\bibfnamefont {R.}~\bibnamefont {Hafner}}, \bibinfo {author} {\bibfnamefont {D.}~\bibnamefont {Spi\ifmmode~\check{s}\else \v{s}\fi{}\'ak}}, \bibinfo {author} {\bibfnamefont {R.}~\bibnamefont {Lorenz}},\ and\ \bibinfo {author} {\bibfnamefont {J.}~\bibnamefont {Hafner}},\ }\bibfield  {title} {\bibinfo {title} {Magnetic ground state of \text{Cr} in density-functional theory},\ }\href {https://doi.org/10.1103/PhysRevB.65.184432} {\bibfield  {journal} {\bibinfo  {journal} {Phys. Rev. B}\ }\textbf {\bibinfo {volume} {65}},\ \bibinfo {pages} {184432} (\bibinfo {year} {2002})}\BibitemShut {NoStop}%
\bibitem [{\citenamefont {Kvashnin}\ \emph {et~al.}(2016)\citenamefont {Kvashnin}, \citenamefont {Cardias}, \citenamefont {Szilva}, \citenamefont {Di~Marco}, \citenamefont {Katsnelson}, \citenamefont {Lichtenstein}, \citenamefont {Nordstr{\"o}m}, \citenamefont {Klautau},\ and\ \citenamefont {Eriksson}}]{kvashnin2016microscopic}%
  \BibitemOpen
  \bibfield  {author} {\bibinfo {author} {\bibfnamefont {Y.~O.}\ \bibnamefont {Kvashnin}}, \bibinfo {author} {\bibfnamefont {R.}~\bibnamefont {Cardias}}, \bibinfo {author} {\bibfnamefont {A.}~\bibnamefont {Szilva}}, \bibinfo {author} {\bibfnamefont {I.}~\bibnamefont {Di~Marco}}, \bibinfo {author} {\bibfnamefont {M.}~\bibnamefont {Katsnelson}}, \bibinfo {author} {\bibfnamefont {A.}~\bibnamefont {Lichtenstein}}, \bibinfo {author} {\bibfnamefont {L.}~\bibnamefont {Nordstr{\"o}m}}, \bibinfo {author} {\bibfnamefont {A.}~\bibnamefont {Klautau}},\ and\ \bibinfo {author} {\bibfnamefont {O.}~\bibnamefont {Eriksson}},\ }\bibfield  {title} {\bibinfo {title} {Microscopic origin of heisenberg and non-heisenberg exchange interactions in ferromagnetic bcc \text{Fe}},\ }\href@noop {} {\bibfield  {journal} {\bibinfo  {journal} {Physical review letters}\ }\textbf {\bibinfo {volume} {116}},\ \bibinfo {pages} {217202} (\bibinfo {year} {2016})}\BibitemShut {NoStop}%
\bibitem [{\citenamefont {Cardias}\ \emph {et~al.}(2017)\citenamefont {Cardias}, \citenamefont {Szilva}, \citenamefont {Bergman}, \citenamefont {Marco}, \citenamefont {Katsnelson}, \citenamefont {Lichtenstein}, \citenamefont {Nordstr{\"o}m}, \citenamefont {Klautau}, \citenamefont {Eriksson},\ and\ \citenamefont {Kvashnin}}]{cardias2017bethe}%
  \BibitemOpen
  \bibfield  {author} {\bibinfo {author} {\bibfnamefont {R.}~\bibnamefont {Cardias}}, \bibinfo {author} {\bibfnamefont {A.}~\bibnamefont {Szilva}}, \bibinfo {author} {\bibfnamefont {A.}~\bibnamefont {Bergman}}, \bibinfo {author} {\bibfnamefont {I.~D.}\ \bibnamefont {Marco}}, \bibinfo {author} {\bibfnamefont {M.}~\bibnamefont {Katsnelson}}, \bibinfo {author} {\bibfnamefont {A.}~\bibnamefont {Lichtenstein}}, \bibinfo {author} {\bibfnamefont {L.}~\bibnamefont {Nordstr{\"o}m}}, \bibinfo {author} {\bibfnamefont {A.}~\bibnamefont {Klautau}}, \bibinfo {author} {\bibfnamefont {O.}~\bibnamefont {Eriksson}},\ and\ \bibinfo {author} {\bibfnamefont {Y.~O.}\ \bibnamefont {Kvashnin}},\ }\bibfield  {title} {\bibinfo {title} {The \text{Bethe-Slater} curve revisited; new insights from electronic structure theory},\ }\href@noop {} {\bibfield  {journal} {\bibinfo  {journal} {Scientific reports}\ }\textbf {\bibinfo {volume} {7}},\ \bibinfo {pages} {4058} (\bibinfo {year} {2017})}\BibitemShut {NoStop}%
\bibitem [{\citenamefont {Zaoui}\ \emph {et~al.}(2020)\citenamefont {Zaoui}, \citenamefont {Bendaoud}, \citenamefont {Obodo}, \citenamefont {Beldi},\ and\ \citenamefont {Bouhafs}}]{zaoui2020competition}%
  \BibitemOpen
  \bibfield  {author} {\bibinfo {author} {\bibfnamefont {Y.}~\bibnamefont {Zaoui}}, \bibinfo {author} {\bibfnamefont {H.}~\bibnamefont {Bendaoud}}, \bibinfo {author} {\bibfnamefont {K.}~\bibnamefont {Obodo}}, \bibinfo {author} {\bibfnamefont {L.}~\bibnamefont {Beldi}},\ and\ \bibinfo {author} {\bibfnamefont {B.}~\bibnamefont {Bouhafs}},\ }\bibfield  {title} {\bibinfo {title} {Competition between the hcp nonmagnetic and antiferromagnetic phases in the transition path of \text{Fe} under pressure},\ }\href@noop {} {\bibfield  {journal} {\bibinfo  {journal} {Journal of Magnetism and Magnetic Materials}\ }\textbf {\bibinfo {volume} {499}},\ \bibinfo {pages} {166312} (\bibinfo {year} {2020})}\BibitemShut {NoStop}%
\bibitem [{\citenamefont {Punkkinen}\ \emph {et~al.}(2011{\natexlab{a}})\citenamefont {Punkkinen}, \citenamefont {Hu}, \citenamefont {Kwon}, \citenamefont {Johansson}, \citenamefont {Koll{\'a}r},\ and\ \citenamefont {Vitos}}]{punkkinen2011surface}%
  \BibitemOpen
  \bibfield  {author} {\bibinfo {author} {\bibfnamefont {M.~P.}\ \bibnamefont {Punkkinen}}, \bibinfo {author} {\bibfnamefont {Q.-M.}\ \bibnamefont {Hu}}, \bibinfo {author} {\bibfnamefont {S.~K.}\ \bibnamefont {Kwon}}, \bibinfo {author} {\bibfnamefont {B.}~\bibnamefont {Johansson}}, \bibinfo {author} {\bibfnamefont {J.}~\bibnamefont {Koll{\'a}r}},\ and\ \bibinfo {author} {\bibfnamefont {L.}~\bibnamefont {Vitos}},\ }\bibfield  {title} {\bibinfo {title} {Surface properties of 3$d$ transition metals},\ }\href@noop {} {\bibfield  {journal} {\bibinfo  {journal} {Philosophical Magazine}\ }\textbf {\bibinfo {volume} {91}},\ \bibinfo {pages} {3627} (\bibinfo {year} {2011}{\natexlab{a}})}\BibitemShut {NoStop}%
\bibitem [{\citenamefont {K{\'a}das}\ \emph {et~al.}(2009)\citenamefont {K{\'a}das}, \citenamefont {Lindquist}, \citenamefont {Eriksson}, \citenamefont {Johansson},\ and\ \citenamefont {Vitos}}]{kadas2009magnetism}%
  \BibitemOpen
  \bibfield  {author} {\bibinfo {author} {\bibfnamefont {K.}~\bibnamefont {K{\'a}das}}, \bibinfo {author} {\bibfnamefont {M.}~\bibnamefont {Lindquist}}, \bibinfo {author} {\bibfnamefont {O.}~\bibnamefont {Eriksson}}, \bibinfo {author} {\bibfnamefont {B.}~\bibnamefont {Johansson}},\ and\ \bibinfo {author} {\bibfnamefont {L.}~\bibnamefont {Vitos}},\ }\bibfield  {title} {\bibinfo {title} {Magnetism-driven anomalous surface alloying between \text{Cu} and \text{Cr}},\ }\href@noop {} {\bibfield  {journal} {\bibinfo  {journal} {Applied Physics Letters}\ }\textbf {\bibinfo {volume} {94}} (\bibinfo {year} {2009})}\BibitemShut {NoStop}%
\bibitem [{\citenamefont {Vitos}\ \emph {et~al.}(2006)\citenamefont {Vitos}, \citenamefont {Korzhavyi},\ and\ \citenamefont {Johansson}}]{vitos2006evidence}%
  \BibitemOpen
  \bibfield  {author} {\bibinfo {author} {\bibfnamefont {L.}~\bibnamefont {Vitos}}, \bibinfo {author} {\bibfnamefont {P.~A.}\ \bibnamefont {Korzhavyi}},\ and\ \bibinfo {author} {\bibfnamefont {B.}~\bibnamefont {Johansson}},\ }\bibfield  {title} {\bibinfo {title} {Evidence of large magnetostructural effects in austenitic stainless steels},\ }\href@noop {} {\bibfield  {journal} {\bibinfo  {journal} {Physical review letters}\ }\textbf {\bibinfo {volume} {96}},\ \bibinfo {pages} {117210} (\bibinfo {year} {2006})}\BibitemShut {NoStop}%
\bibitem [{\citenamefont {Li}\ \emph {et~al.}(2016)\citenamefont {Li}, \citenamefont {Lu}, \citenamefont {Kim}, \citenamefont {Kokko}, \citenamefont {Hertzman}, \citenamefont {Kwon},\ and\ \citenamefont {Vitos}}]{li2016first}%
  \BibitemOpen
  \bibfield  {author} {\bibinfo {author} {\bibfnamefont {W.}~\bibnamefont {Li}}, \bibinfo {author} {\bibfnamefont {S.}~\bibnamefont {Lu}}, \bibinfo {author} {\bibfnamefont {D.}~\bibnamefont {Kim}}, \bibinfo {author} {\bibfnamefont {K.}~\bibnamefont {Kokko}}, \bibinfo {author} {\bibfnamefont {S.}~\bibnamefont {Hertzman}}, \bibinfo {author} {\bibfnamefont {S.~K.}\ \bibnamefont {Kwon}},\ and\ \bibinfo {author} {\bibfnamefont {L.}~\bibnamefont {Vitos}},\ }\bibfield  {title} {\bibinfo {title} {First-principles prediction of the deformation modes in austenitic \text{Fe-Cr-Ni} alloys},\ }\href@noop {} {\bibfield  {journal} {\bibinfo  {journal} {Applied Physics Letters}\ }\textbf {\bibinfo {volume} {108}} (\bibinfo {year} {2016})}\BibitemShut {NoStop}%
\bibitem [{\citenamefont {Zelen\'y}\ \emph {et~al.}(2011)\citenamefont {Zelen\'y}, \citenamefont {Fri\'ak},\ and\ \citenamefont {\ifmmode~\check{S}\else \v{S}\fi{}ob}}]{PhysRevB.83.184424}%
  \BibitemOpen
  \bibfield  {author} {\bibinfo {author} {\bibfnamefont {M.}~\bibnamefont {Zelen\'y}}, \bibinfo {author} {\bibfnamefont {M.}~\bibnamefont {Fri\'ak}},\ and\ \bibinfo {author} {\bibfnamefont {M.}~\bibnamefont {\ifmmode~\check{S}\else \v{S}\fi{}ob}},\ }\bibfield  {title} {\bibinfo {title} {$ab$ $initio$ study of energetics and magnetism of \text{Fe}, \text{Co}, and \text{Ni} along the trigonal deformation path},\ }\href {https://doi.org/10.1103/PhysRevB.83.184424} {\bibfield  {journal} {\bibinfo  {journal} {Phys. Rev. B}\ }\textbf {\bibinfo {volume} {83}},\ \bibinfo {pages} {184424} (\bibinfo {year} {2011})}\BibitemShut {NoStop}%
\bibitem [{\citenamefont {Shirane}\ \emph {et~al.}(1964)\citenamefont {Shirane}, \citenamefont {Nathans},\ and\ \citenamefont {Chen}}]{PhysRev.134.A1547}%
  \BibitemOpen
  \bibfield  {author} {\bibinfo {author} {\bibfnamefont {G.}~\bibnamefont {Shirane}}, \bibinfo {author} {\bibfnamefont {R.}~\bibnamefont {Nathans}},\ and\ \bibinfo {author} {\bibfnamefont {C.~W.}\ \bibnamefont {Chen}},\ }\bibfield  {title} {\bibinfo {title} {Magnetic moments and unpaired spin densities in the \text{Fe-Rh} alloys},\ }\href {https://doi.org/10.1103/PhysRev.134.A1547} {\bibfield  {journal} {\bibinfo  {journal} {Phys. Rev.}\ }\textbf {\bibinfo {volume} {134}},\ \bibinfo {pages} {A1547} (\bibinfo {year} {1964})}\BibitemShut {NoStop}%
\bibitem [{\citenamefont {Shirane}\ \emph {et~al.}(1963)\citenamefont {Shirane}, \citenamefont {Chen}, \citenamefont {Flinn},\ and\ \citenamefont {Nathans}}]{PhysRev.131.183}%
  \BibitemOpen
  \bibfield  {author} {\bibinfo {author} {\bibfnamefont {G.}~\bibnamefont {Shirane}}, \bibinfo {author} {\bibfnamefont {C.~W.}\ \bibnamefont {Chen}}, \bibinfo {author} {\bibfnamefont {P.~A.}\ \bibnamefont {Flinn}},\ and\ \bibinfo {author} {\bibfnamefont {R.}~\bibnamefont {Nathans}},\ }\bibfield  {title} {\bibinfo {title} {M\"ossbauer study of hyperfine fields and isomer shifts in the \text{Fe-Rh} alloys},\ }\href {https://doi.org/10.1103/PhysRev.131.183} {\bibfield  {journal} {\bibinfo  {journal} {Phys. Rev.}\ }\textbf {\bibinfo {volume} {131}},\ \bibinfo {pages} {183} (\bibinfo {year} {1963})}\BibitemShut {NoStop}%
\bibitem [{\citenamefont {Zarkevich}\ and\ \citenamefont {Johnson}(2018)}]{zarkevich2018ferh}%
  \BibitemOpen
  \bibfield  {author} {\bibinfo {author} {\bibfnamefont {N.~A.}\ \bibnamefont {Zarkevich}}\ and\ \bibinfo {author} {\bibfnamefont {D.~D.}\ \bibnamefont {Johnson}},\ }\bibfield  {title} {\bibinfo {title} {\text{FeRh} ground state and martensitic transformation},\ }\href@noop {} {\bibfield  {journal} {\bibinfo  {journal} {Physical Review B}\ }\textbf {\bibinfo {volume} {97}},\ \bibinfo {pages} {014202} (\bibinfo {year} {2018})}\BibitemShut {NoStop}%
\bibitem [{\citenamefont {Vieira}\ \emph {et~al.}(2021)\citenamefont {Vieira}, \citenamefont {Eriksson}, \citenamefont {Bergman},\ and\ \citenamefont {Herper}}]{VIEIRA2021157811}%
  \BibitemOpen
  \bibfield  {author} {\bibinfo {author} {\bibfnamefont {R.~M.}\ \bibnamefont {Vieira}}, \bibinfo {author} {\bibfnamefont {O.}~\bibnamefont {Eriksson}}, \bibinfo {author} {\bibfnamefont {A.}~\bibnamefont {Bergman}},\ and\ \bibinfo {author} {\bibfnamefont {H.}~\bibnamefont {Herper}},\ }\bibfield  {title} {\bibinfo {title} {High-throughput compatible approach for entropy estimation in magnetocaloric materials: \text{FeRh} as a test case},\ }\href {https://doi.org/https://doi.org/10.1016/j.jallcom.2020.157811} {\bibfield  {journal} {\bibinfo  {journal} {Journal of Alloys and Compounds}\ }\textbf {\bibinfo {volume} {857}},\ \bibinfo {pages} {157811} (\bibinfo {year} {2021})}\BibitemShut {NoStop}%
\bibitem [{\citenamefont {Wolloch}\ \emph {et~al.}(2016)\citenamefont {Wolloch}, \citenamefont {Gruner}, \citenamefont {Keune}, \citenamefont {Mohn}, \citenamefont {Redinger}, \citenamefont {Hofer}, \citenamefont {Suess}, \citenamefont {Podloucky}, \citenamefont {Landers}, \citenamefont {Salamon}, \citenamefont {Scheibel}, \citenamefont {Spoddig}, \citenamefont {Witte}, \citenamefont {Roldan~Cuenya}, \citenamefont {Gutfleisch}, \citenamefont {Hu}, \citenamefont {Zhao}, \citenamefont {Toellner}, \citenamefont {Alp}, \citenamefont {Siewert}, \citenamefont {Entel}, \citenamefont {Pentcheva},\ and\ \citenamefont {Wende}}]{PhysRevB.94.174435}%
  \BibitemOpen
  \bibfield  {author} {\bibinfo {author} {\bibfnamefont {M.}~\bibnamefont {Wolloch}}, \bibinfo {author} {\bibfnamefont {M.~E.}\ \bibnamefont {Gruner}}, \bibinfo {author} {\bibfnamefont {W.}~\bibnamefont {Keune}}, \bibinfo {author} {\bibfnamefont {P.}~\bibnamefont {Mohn}}, \bibinfo {author} {\bibfnamefont {J.}~\bibnamefont {Redinger}}, \bibinfo {author} {\bibfnamefont {F.}~\bibnamefont {Hofer}}, \bibinfo {author} {\bibfnamefont {D.}~\bibnamefont {Suess}}, \bibinfo {author} {\bibfnamefont {R.}~\bibnamefont {Podloucky}}, \bibinfo {author} {\bibfnamefont {J.}~\bibnamefont {Landers}}, \bibinfo {author} {\bibfnamefont {S.}~\bibnamefont {Salamon}}, \bibinfo {author} {\bibfnamefont {F.}~\bibnamefont {Scheibel}}, \bibinfo {author} {\bibfnamefont {D.}~\bibnamefont {Spoddig}}, \bibinfo {author} {\bibfnamefont {R.}~\bibnamefont {Witte}}, \bibinfo {author} {\bibfnamefont {B.}~\bibnamefont {Roldan~Cuenya}}, \bibinfo {author} {\bibfnamefont {O.}~\bibnamefont {Gutfleisch}}, \bibinfo {author} {\bibfnamefont {M.~Y.}\
  \bibnamefont {Hu}}, \bibinfo {author} {\bibfnamefont {J.}~\bibnamefont {Zhao}}, \bibinfo {author} {\bibfnamefont {T.}~\bibnamefont {Toellner}}, \bibinfo {author} {\bibfnamefont {E.~E.}\ \bibnamefont {Alp}}, \bibinfo {author} {\bibfnamefont {M.}~\bibnamefont {Siewert}}, \bibinfo {author} {\bibfnamefont {P.}~\bibnamefont {Entel}}, \bibinfo {author} {\bibfnamefont {R.}~\bibnamefont {Pentcheva}},\ and\ \bibinfo {author} {\bibfnamefont {H.}~\bibnamefont {Wende}},\ }\bibfield  {title} {\bibinfo {title} {Impact of lattice dynamics on the phase stability of metamagnetic \text{FeRh}: Bulk and thin films},\ }\href {https://doi.org/10.1103/PhysRevB.94.174435} {\bibfield  {journal} {\bibinfo  {journal} {Phys. Rev. B}\ }\textbf {\bibinfo {volume} {94}},\ \bibinfo {pages} {174435} (\bibinfo {year} {2016})}\BibitemShut {NoStop}%
\bibitem [{\citenamefont {Sánchez-Valdés}\ \emph {et~al.}(2020)\citenamefont {Sánchez-Valdés}, \citenamefont {Gimaev}, \citenamefont {López-Cruz}, \citenamefont {{Sánchez Llamazares}}, \citenamefont {Zverev}, \citenamefont {Tishin}, \citenamefont {Carvalho}, \citenamefont {Aguiar}, \citenamefont {Mudryk},\ and\ \citenamefont {Pecharsky}}]{SANCHEZVALDES2020166130}%
  \BibitemOpen
  \bibfield  {author} {\bibinfo {author} {\bibfnamefont {C.}~\bibnamefont {Sánchez-Valdés}}, \bibinfo {author} {\bibfnamefont {R.}~\bibnamefont {Gimaev}}, \bibinfo {author} {\bibfnamefont {M.}~\bibnamefont {López-Cruz}}, \bibinfo {author} {\bibfnamefont {J.}~\bibnamefont {{Sánchez Llamazares}}}, \bibinfo {author} {\bibfnamefont {V.}~\bibnamefont {Zverev}}, \bibinfo {author} {\bibfnamefont {A.}~\bibnamefont {Tishin}}, \bibinfo {author} {\bibfnamefont {A.}~\bibnamefont {Carvalho}}, \bibinfo {author} {\bibfnamefont {D.}~\bibnamefont {Aguiar}}, \bibinfo {author} {\bibfnamefont {Y.}~\bibnamefont {Mudryk}},\ and\ \bibinfo {author} {\bibfnamefont {V.}~\bibnamefont {Pecharsky}},\ }\bibfield  {title} {\bibinfo {title} {The effect of cooling rate on magnetothermal properties of {$\mathrm{Fe_{49}Rh_{51}}$}},\ }\href {https://doi.org/https://doi.org/10.1016/j.jmmm.2019.166130} {\bibfield  {journal} {\bibinfo  {journal} {Journal of Magnetism and Magnetic Materials}\ }\textbf {\bibinfo {volume} {498}},\ \bibinfo
  {pages} {166130} (\bibinfo {year} {2020})}\BibitemShut {NoStop}%
\bibitem [{\citenamefont {Dastanpour}\ \emph {et~al.}(2024)\citenamefont {Dastanpour}, \citenamefont {Huang}, \citenamefont {Ström}, \citenamefont {Varga}, \citenamefont {Vitos},\ and\ \citenamefont {Schönecker}}]{DASTANPOUR2024173977}%
  \BibitemOpen
  \bibfield  {author} {\bibinfo {author} {\bibfnamefont {E.}~\bibnamefont {Dastanpour}}, \bibinfo {author} {\bibfnamefont {S.}~\bibnamefont {Huang}}, \bibinfo {author} {\bibfnamefont {V.}~\bibnamefont {Ström}}, \bibinfo {author} {\bibfnamefont {L.~K.}\ \bibnamefont {Varga}}, \bibinfo {author} {\bibfnamefont {L.}~\bibnamefont {Vitos}},\ and\ \bibinfo {author} {\bibfnamefont {S.}~\bibnamefont {Schönecker}},\ }\bibfield  {title} {\bibinfo {title} {An assessment of the {$\mathrm{Al_{50} Cr_{21-x} Mn_{17+x} Co_{12} (x=0, 4, 8)}$} high-entropy alloys for magnetocaloric refrigeration application},\ }\href {https://doi.org/https://doi.org/10.1016/j.jallcom.2024.173977} {\bibfield  {journal} {\bibinfo  {journal} {Journal of Alloys and Compounds}\ }\textbf {\bibinfo {volume} {984}},\ \bibinfo {pages} {173977} (\bibinfo {year} {2024})}\BibitemShut {NoStop}%
\bibitem [{\citenamefont {Dastanpour}\ \emph {et~al.}(2022)\citenamefont {Dastanpour}, \citenamefont {Huang}, \citenamefont {Sch{\"o}necker}, \citenamefont {Mao}, \citenamefont {Str{\"o}m}, \citenamefont {Eriksson}, \citenamefont {Varga},\ and\ \citenamefont {Vitos}}]{dastanpour2022structural}%
  \BibitemOpen
  \bibfield  {author} {\bibinfo {author} {\bibfnamefont {E.}~\bibnamefont {Dastanpour}}, \bibinfo {author} {\bibfnamefont {S.}~\bibnamefont {Huang}}, \bibinfo {author} {\bibfnamefont {S.}~\bibnamefont {Sch{\"o}necker}}, \bibinfo {author} {\bibfnamefont {H.}~\bibnamefont {Mao}}, \bibinfo {author} {\bibfnamefont {V.}~\bibnamefont {Str{\"o}m}}, \bibinfo {author} {\bibfnamefont {O.}~\bibnamefont {Eriksson}}, \bibinfo {author} {\bibfnamefont {L.~K.}\ \bibnamefont {Varga}},\ and\ \bibinfo {author} {\bibfnamefont {L.}~\bibnamefont {Vitos}},\ }\bibfield  {title} {\bibinfo {title} {On the structural and magnetic properties of \text{Al-rich} high entropy alloys: a joint experimental-theoretical study},\ }\href@noop {} {\bibfield  {journal} {\bibinfo  {journal} {Journal of Physics D: Applied Physics}\ }\textbf {\bibinfo {volume} {56}},\ \bibinfo {pages} {015003} (\bibinfo {year} {2022})}\BibitemShut {NoStop}%
\bibitem [{\citenamefont {Huang}\ \emph {et~al.}(2023)\citenamefont {Huang}, \citenamefont {Dastanpour}, \citenamefont {Sch{\"o}necker}, \citenamefont {Str{\"o}m}, \citenamefont {Chai}, \citenamefont {Kiss}, \citenamefont {Varga}, \citenamefont {Jin}, \citenamefont {Eriksson},\ and\ \citenamefont {Vitos}}]{huang2023combinatorial}%
  \BibitemOpen
  \bibfield  {author} {\bibinfo {author} {\bibfnamefont {S.}~\bibnamefont {Huang}}, \bibinfo {author} {\bibfnamefont {E.}~\bibnamefont {Dastanpour}}, \bibinfo {author} {\bibfnamefont {S.}~\bibnamefont {Sch{\"o}necker}}, \bibinfo {author} {\bibfnamefont {V.}~\bibnamefont {Str{\"o}m}}, \bibinfo {author} {\bibfnamefont {G.}~\bibnamefont {Chai}}, \bibinfo {author} {\bibfnamefont {L.~F.}\ \bibnamefont {Kiss}}, \bibinfo {author} {\bibfnamefont {L.~K.}\ \bibnamefont {Varga}}, \bibinfo {author} {\bibfnamefont {H.}~\bibnamefont {Jin}}, \bibinfo {author} {\bibfnamefont {O.}~\bibnamefont {Eriksson}},\ and\ \bibinfo {author} {\bibfnamefont {L.}~\bibnamefont {Vitos}},\ }\bibfield  {title} {\bibinfo {title} {Combinatorial design of partial ordered \text{Al-Cr-Mn-Co} medium-entropy alloys for room temperature magnetic refrigeration applications},\ }\href@noop {} {\bibfield  {journal} {\bibinfo  {journal} {Applied Physics Letters}\ }\textbf {\bibinfo {volume} {123}},\ \bibinfo {pages} {044103} (\bibinfo {year}
  {2023})}\BibitemShut {NoStop}%
\bibitem [{\citenamefont {Dastanpour}\ \emph {et~al.}(2025{\natexlab{a}})\citenamefont {Dastanpour}, \citenamefont {Aihemaiti}, \citenamefont {Huang}, \citenamefont {Str{\"o}m}, \citenamefont {Varga},\ and\ \citenamefont {Vitos}}]{dastanpour2025structural}%
  \BibitemOpen
  \bibfield  {author} {\bibinfo {author} {\bibfnamefont {E.}~\bibnamefont {Dastanpour}}, \bibinfo {author} {\bibfnamefont {H.}~\bibnamefont {Aihemaiti}}, \bibinfo {author} {\bibfnamefont {S.}~\bibnamefont {Huang}}, \bibinfo {author} {\bibfnamefont {V.}~\bibnamefont {Str{\"o}m}}, \bibinfo {author} {\bibfnamefont {L.~K.}\ \bibnamefont {Varga}},\ and\ \bibinfo {author} {\bibfnamefont {L.}~\bibnamefont {Vitos}},\ }\bibfield  {title} {\bibinfo {title} {Structural and ferromagnetic response of \text{B2}-type {$\mathrm{Al_{45} Mn_{41.8} X_{13.2} (X=Fe, Co, Ni)}$} alloys},\ }\href@noop {} {\bibfield  {journal} {\bibinfo  {journal} {Magnetochemistry}\ }\textbf {\bibinfo {volume} {11}},\ \bibinfo {pages} {67} (\bibinfo {year} {2025}{\natexlab{a}})}\BibitemShut {NoStop}%
\bibitem [{\citenamefont {Kohn}\ and\ \citenamefont {Sham}(1965)}]{PhysRev.140.A1133}%
  \BibitemOpen
  \bibfield  {author} {\bibinfo {author} {\bibfnamefont {W.}~\bibnamefont {Kohn}}\ and\ \bibinfo {author} {\bibfnamefont {L.~J.}\ \bibnamefont {Sham}},\ }\bibfield  {title} {\bibinfo {title} {Self-consistent equations including exchange and correlation effects},\ }\href {https://doi.org/10.1103/PhysRev.140.A1133} {\bibfield  {journal} {\bibinfo  {journal} {Phys. Rev.}\ }\textbf {\bibinfo {volume} {140}},\ \bibinfo {pages} {A1133} (\bibinfo {year} {1965})}\BibitemShut {NoStop}%
\bibitem [{\citenamefont {Perdew}\ and\ \citenamefont {Wang}(1992)}]{PhysRevB.45.13244}%
  \BibitemOpen
  \bibfield  {author} {\bibinfo {author} {\bibfnamefont {J.~P.}\ \bibnamefont {Perdew}}\ and\ \bibinfo {author} {\bibfnamefont {Y.}~\bibnamefont {Wang}},\ }\bibfield  {title} {\bibinfo {title} {Accurate and simple analytic representation of the electron-gas correlation energy},\ }\href {https://doi.org/10.1103/PhysRevB.45.13244} {\bibfield  {journal} {\bibinfo  {journal} {Phys. Rev. B}\ }\textbf {\bibinfo {volume} {45}},\ \bibinfo {pages} {13244} (\bibinfo {year} {1992})}\BibitemShut {NoStop}%
\bibitem [{\citenamefont {Vitos}(2001)}]{PhysRevB.64.014107}%
  \BibitemOpen
  \bibfield  {author} {\bibinfo {author} {\bibfnamefont {L.}~\bibnamefont {Vitos}},\ }\bibfield  {title} {\bibinfo {title} {Total-energy method based on the exact muffin-tin orbitals theory},\ }\href@noop {} {\bibfield  {journal} {\bibinfo  {journal} {Phys. Rev. B}\ }\textbf {\bibinfo {volume} {64}},\ \bibinfo {pages} {014107} (\bibinfo {year} {2001})}\BibitemShut {NoStop}%
\bibitem [{\citenamefont {Vitos}(2007)}]{vitos2007computational}%
  \BibitemOpen
  \bibfield  {author} {\bibinfo {author} {\bibfnamefont {L.}~\bibnamefont {Vitos}},\ }\href@noop {} {\emph {\bibinfo {title} {Computational quantum mechanics for materials engineers: the EMTO method and applications}}}\ (\bibinfo  {publisher} {Springer Science \& Business Media},\ \bibinfo {year} {2007})\BibitemShut {NoStop}%
\bibitem [{\citenamefont {Soven}(1967)}]{PhysRev.156.809}%
  \BibitemOpen
  \bibfield  {author} {\bibinfo {author} {\bibfnamefont {P.}~\bibnamefont {Soven}},\ }\bibfield  {title} {\bibinfo {title} {Coherent-potential model of substitutional disordered alloys},\ }\href@noop {} {\bibfield  {journal} {\bibinfo  {journal} {Phys. Rev.}\ }\textbf {\bibinfo {volume} {156}},\ \bibinfo {pages} {809} (\bibinfo {year} {1967})}\BibitemShut {NoStop}%
\bibitem [{\citenamefont {Gyorffy}(1972)}]{PhysRevB.5.2382}%
  \BibitemOpen
  \bibfield  {author} {\bibinfo {author} {\bibfnamefont {B.~L.}\ \bibnamefont {Gyorffy}},\ }\bibfield  {title} {\bibinfo {title} {Coherent-potential approximation for a nonoverlapping-muffin-tin-potential model of random substitutional alloys},\ }\href@noop {} {\bibfield  {journal} {\bibinfo  {journal} {Phys. Rev. B}\ }\textbf {\bibinfo {volume} {5}},\ \bibinfo {pages} {2382} (\bibinfo {year} {1972})}\BibitemShut {NoStop}%
\bibitem [{\citenamefont {Vitos}\ \emph {et~al.}(2001)\citenamefont {Vitos}, \citenamefont {Abrikosov},\ and\ \citenamefont {Johansson}}]{PhysRevLett.87.156401}%
  \BibitemOpen
  \bibfield  {author} {\bibinfo {author} {\bibfnamefont {L.}~\bibnamefont {Vitos}}, \bibinfo {author} {\bibfnamefont {I.~A.}\ \bibnamefont {Abrikosov}},\ and\ \bibinfo {author} {\bibfnamefont {B.}~\bibnamefont {Johansson}},\ }\bibfield  {title} {\bibinfo {title} {Anisotropic lattice distortions in random alloys from first-principles theory},\ }\href@noop {} {\bibfield  {journal} {\bibinfo  {journal} {Phys. Rev. Lett.}\ }\textbf {\bibinfo {volume} {87}},\ \bibinfo {pages} {156401} (\bibinfo {year} {2001})}\BibitemShut {NoStop}%
\bibitem [{\citenamefont {Vitos}\ \emph {et~al.}(1997)\citenamefont {Vitos}, \citenamefont {Koll{\'a}r},\ and\ \citenamefont {Skriver}}]{vitos1997full}%
  \BibitemOpen
  \bibfield  {author} {\bibinfo {author} {\bibfnamefont {L.}~\bibnamefont {Vitos}}, \bibinfo {author} {\bibfnamefont {J.}~\bibnamefont {Koll{\'a}r}},\ and\ \bibinfo {author} {\bibfnamefont {H.~L.}\ \bibnamefont {Skriver}},\ }\bibfield  {title} {\bibinfo {title} {Full charge-density scheme with a kinetic-energy correction: Application to ground-state properties of the 4$d$ metals},\ }\href@noop {} {\bibfield  {journal} {\bibinfo  {journal} {Physical Review B}\ }\textbf {\bibinfo {volume} {55}},\ \bibinfo {pages} {13521} (\bibinfo {year} {1997})}\BibitemShut {NoStop}%
\bibitem [{\citenamefont {Koll{\'a}r}\ \emph {et~al.}(2000)\citenamefont {Koll{\'a}r}, \citenamefont {Vitos}, \citenamefont {Skriver},\ and\ \citenamefont {Dreyss{\'e}}}]{kollar2000electronic}%
  \BibitemOpen
  \bibfield  {author} {\bibinfo {author} {\bibfnamefont {J.}~\bibnamefont {Koll{\'a}r}}, \bibinfo {author} {\bibfnamefont {L.}~\bibnamefont {Vitos}}, \bibinfo {author} {\bibfnamefont {H.}~\bibnamefont {Skriver}},\ and\ \bibinfo {author} {\bibfnamefont {H.}~\bibnamefont {Dreyss{\'e}}},\ }\bibfield  {title} {\bibinfo {title} {Electronic structure and physical properties of solids: the uses of the \text{LMTO} method},\ }\href@noop {} {\bibfield  {journal} {\bibinfo  {journal} {Lecture Notes in Physics}\ ,\ \bibinfo {pages} {85}} (\bibinfo {year} {2000})}\BibitemShut {NoStop}%
\bibitem [{\citenamefont {Kissavos}\ \emph {et~al.}(2006)\citenamefont {Kissavos}, \citenamefont {Simak}, \citenamefont {Olsson}, \citenamefont {Vitos},\ and\ \citenamefont {Abrikosov}}]{kissavos2006total}%
  \BibitemOpen
  \bibfield  {author} {\bibinfo {author} {\bibfnamefont {A.~E.}\ \bibnamefont {Kissavos}}, \bibinfo {author} {\bibfnamefont {S.~I.}\ \bibnamefont {Simak}}, \bibinfo {author} {\bibfnamefont {P.}~\bibnamefont {Olsson}}, \bibinfo {author} {\bibfnamefont {L.}~\bibnamefont {Vitos}},\ and\ \bibinfo {author} {\bibfnamefont {I.~A.}\ \bibnamefont {Abrikosov}},\ }\bibfield  {title} {\bibinfo {title} {Total energy calculations for systems with magnetic and chemical disorder},\ }\href@noop {} {\bibfield  {journal} {\bibinfo  {journal} {Computational materials science}\ }\textbf {\bibinfo {volume} {35}},\ \bibinfo {pages} {1} (\bibinfo {year} {2006})}\BibitemShut {NoStop}%
\bibitem [{\citenamefont {Perdew}\ \emph {et~al.}(1996)\citenamefont {Perdew}, \citenamefont {Burke},\ and\ \citenamefont {Ernzerhof}}]{PhysRevLett.77.3865}%
  \BibitemOpen
  \bibfield  {author} {\bibinfo {author} {\bibfnamefont {J.~P.}\ \bibnamefont {Perdew}}, \bibinfo {author} {\bibfnamefont {K.}~\bibnamefont {Burke}},\ and\ \bibinfo {author} {\bibfnamefont {M.}~\bibnamefont {Ernzerhof}},\ }\bibfield  {title} {\bibinfo {title} {Generalized gradient approximation made simple},\ }\href@noop {} {\bibfield  {journal} {\bibinfo  {journal} {Phys. Rev. Lett.}\ }\textbf {\bibinfo {volume} {77}},\ \bibinfo {pages} {3865} (\bibinfo {year} {1996})}\BibitemShut {NoStop}%
\bibitem [{\citenamefont {Gyorffy}\ \emph {et~al.}(1985)\citenamefont {Gyorffy}, \citenamefont {Pindor}, \citenamefont {Staunton}, \citenamefont {Stocks},\ and\ \citenamefont {Winter}}]{gyorffy1985first}%
  \BibitemOpen
  \bibfield  {author} {\bibinfo {author} {\bibfnamefont {B.}~\bibnamefont {Gyorffy}}, \bibinfo {author} {\bibfnamefont {A.}~\bibnamefont {Pindor}}, \bibinfo {author} {\bibfnamefont {J.}~\bibnamefont {Staunton}}, \bibinfo {author} {\bibfnamefont {G.}~\bibnamefont {Stocks}},\ and\ \bibinfo {author} {\bibfnamefont {H.}~\bibnamefont {Winter}},\ }\bibfield  {title} {\bibinfo {title} {A first-principles theory of ferromagnetic phase transitions in metals},\ }\href@noop {} {\bibfield  {journal} {\bibinfo  {journal} {Journal of Physics F: Metal Physics}\ }\textbf {\bibinfo {volume} {15}},\ \bibinfo {pages} {1337} (\bibinfo {year} {1985})}\BibitemShut {NoStop}%
\bibitem [{\citenamefont {Ruban}\ and\ \citenamefont {Dehghani}(2016)}]{ruban2016atomic}%
  \BibitemOpen
  \bibfield  {author} {\bibinfo {author} {\bibfnamefont {A.~V.}\ \bibnamefont {Ruban}}\ and\ \bibinfo {author} {\bibfnamefont {M.}~\bibnamefont {Dehghani}},\ }\bibfield  {title} {\bibinfo {title} {Atomic configuration and properties of austenitic steels at finite temperature: Effect of longitudinal spin fluctuations},\ }\href@noop {} {\bibfield  {journal} {\bibinfo  {journal} {Physical Review B}\ }\textbf {\bibinfo {volume} {94}},\ \bibinfo {pages} {104111} (\bibinfo {year} {2016})}\BibitemShut {NoStop}%
\bibitem [{\citenamefont {Moruzzi}\ \emph {et~al.}(1988)\citenamefont {Moruzzi}, \citenamefont {Janak},\ and\ \citenamefont {Schwarz}}]{moruzzi1988calculated}%
  \BibitemOpen
  \bibfield  {author} {\bibinfo {author} {\bibfnamefont {V.}~\bibnamefont {Moruzzi}}, \bibinfo {author} {\bibfnamefont {J.}~\bibnamefont {Janak}},\ and\ \bibinfo {author} {\bibfnamefont {K.}~\bibnamefont {Schwarz}},\ }\bibfield  {title} {\bibinfo {title} {Calculated thermal properties of metals},\ }\href@noop {} {\bibfield  {journal} {\bibinfo  {journal} {Physical Review B}\ }\textbf {\bibinfo {volume} {37}},\ \bibinfo {pages} {790} (\bibinfo {year} {1988})}\BibitemShut {NoStop}%
\bibitem [{\citenamefont {Vitos}\ \emph {et~al.}(2000)\citenamefont {Vitos}, \citenamefont {Skriver}, \citenamefont {Johansson},\ and\ \citenamefont {Koll{\'a}r}}]{vitos2000application}%
  \BibitemOpen
  \bibfield  {author} {\bibinfo {author} {\bibfnamefont {L.}~\bibnamefont {Vitos}}, \bibinfo {author} {\bibfnamefont {H.~L.}\ \bibnamefont {Skriver}}, \bibinfo {author} {\bibfnamefont {B.}~\bibnamefont {Johansson}},\ and\ \bibinfo {author} {\bibfnamefont {J.}~\bibnamefont {Koll{\'a}r}},\ }\bibfield  {title} {\bibinfo {title} {Application of the exact muffin-tin orbitals theory: the spherical cell approximation},\ }\href@noop {} {\bibfield  {journal} {\bibinfo  {journal} {Computational materials science}\ }\textbf {\bibinfo {volume} {18}},\ \bibinfo {pages} {24} (\bibinfo {year} {2000})}\BibitemShut {NoStop}%
\bibitem [{\citenamefont {Eriksson}\ \emph {et~al.}(2017)\citenamefont {Eriksson}, \citenamefont {Bergman}, \citenamefont {Bergqvist},\ and\ \citenamefont {Hellsvik}}]{Eriksson2017}%
  \BibitemOpen
  \bibfield  {author} {\bibinfo {author} {\bibfnamefont {O.}~\bibnamefont {Eriksson}}, \bibinfo {author} {\bibfnamefont {A.}~\bibnamefont {Bergman}}, \bibinfo {author} {\bibfnamefont {L.}~\bibnamefont {Bergqvist}},\ and\ \bibinfo {author} {\bibfnamefont {A.~J.}\ \bibnamefont {Hellsvik}},\ }\href {https://github.com/UppASD/UppASD} {\emph {\bibinfo {title} {Atomistic Spin Dynamics: Foundations and Applications}}}\ (\bibinfo  {publisher} {Oxford University Press},\ \bibinfo {address} {Oxford, New York},\ \bibinfo {year} {2017})\BibitemShut {NoStop}%
\bibitem [{\citenamefont {Punkkinen}\ \emph {et~al.}(2011{\natexlab{b}})\citenamefont {Punkkinen}, \citenamefont {Kwon}, \citenamefont {Koll{\'a}r}, \citenamefont {Johansson},\ and\ \citenamefont {Vitos}}]{punkkinen2011compressive}%
  \BibitemOpen
  \bibfield  {author} {\bibinfo {author} {\bibfnamefont {M.~P.~J.}\ \bibnamefont {Punkkinen}}, \bibinfo {author} {\bibfnamefont {S.}~\bibnamefont {Kwon}}, \bibinfo {author} {\bibfnamefont {J.}~\bibnamefont {Koll{\'a}r}}, \bibinfo {author} {\bibfnamefont {B.}~\bibnamefont {Johansson}},\ and\ \bibinfo {author} {\bibfnamefont {L.}~\bibnamefont {Vitos}},\ }\bibfield  {title} {\bibinfo {title} {Compressive surface stress in magnetic transition metals},\ }\href@noop {} {\bibfield  {journal} {\bibinfo  {journal} {Physical Review Letters}\ }\textbf {\bibinfo {volume} {106}},\ \bibinfo {pages} {057202} (\bibinfo {year} {2011}{\natexlab{b}})}\BibitemShut {NoStop}%
\bibitem [{\citenamefont {Kulikov}\ \emph {et~al.}(1999)\citenamefont {Kulikov}, \citenamefont {Postnikov}, \citenamefont {Borstel},\ and\ \citenamefont {Braun}}]{kulikov1999onset}%
  \BibitemOpen
  \bibfield  {author} {\bibinfo {author} {\bibfnamefont {N.}~\bibnamefont {Kulikov}}, \bibinfo {author} {\bibfnamefont {A.}~\bibnamefont {Postnikov}}, \bibinfo {author} {\bibfnamefont {G.}~\bibnamefont {Borstel}},\ and\ \bibinfo {author} {\bibfnamefont {J.}~\bibnamefont {Braun}},\ }\bibfield  {title} {\bibinfo {title} {Onset of magnetism in \text{B2} transition-metal aluminides},\ }\href@noop {} {\bibfield  {journal} {\bibinfo  {journal} {Physical Review B}\ }\textbf {\bibinfo {volume} {59}},\ \bibinfo {pages} {6824} (\bibinfo {year} {1999})}\BibitemShut {NoStop}%
\bibitem [{\citenamefont {Huang}\ \emph {et~al.}(2018{\natexlab{a}})\citenamefont {Huang}, \citenamefont {Li}, \citenamefont {Huang}, \citenamefont {Holmstr{\"o}m},\ and\ \citenamefont {Vitos}}]{huang2018mechanical}%
  \BibitemOpen
  \bibfield  {author} {\bibinfo {author} {\bibfnamefont {S.}~\bibnamefont {Huang}}, \bibinfo {author} {\bibfnamefont {X.}~\bibnamefont {Li}}, \bibinfo {author} {\bibfnamefont {H.}~\bibnamefont {Huang}}, \bibinfo {author} {\bibfnamefont {E.}~\bibnamefont {Holmstr{\"o}m}},\ and\ \bibinfo {author} {\bibfnamefont {L.}~\bibnamefont {Vitos}},\ }\bibfield  {title} {\bibinfo {title} {Mechanical performance of {$\mathrm{FeCrCoMnAl_{x}}$} high-entropy alloys from first-principle},\ }\href@noop {} {\bibfield  {journal} {\bibinfo  {journal} {Materials Chemistry and Physics}\ }\textbf {\bibinfo {volume} {210}},\ \bibinfo {pages} {37} (\bibinfo {year} {2018}{\natexlab{a}})}\BibitemShut {NoStop}%
\bibitem [{\citenamefont {Helander}\ and\ \citenamefont {Tolochko}(1999)}]{helander1999experimental}%
  \BibitemOpen
  \bibfield  {author} {\bibinfo {author} {\bibfnamefont {T.}~\bibnamefont {Helander}}\ and\ \bibinfo {author} {\bibfnamefont {O.}~\bibnamefont {Tolochko}},\ }\bibfield  {title} {\bibinfo {title} {An experimental investigation of possible \text{B2-ordering} in the \text{Al-Cr} system},\ }\href@noop {} {\bibfield  {journal} {\bibinfo  {journal} {Journal of phase equilibria}\ }\textbf {\bibinfo {volume} {20}},\ \bibinfo {pages} {57} (\bibinfo {year} {1999})}\BibitemShut {NoStop}%
\bibitem [{\citenamefont {Mohn}\ \emph {et~al.}(2001)\citenamefont {Mohn}, \citenamefont {Persson}, \citenamefont {Blaha}, \citenamefont {Schwarz}, \citenamefont {Nov{\'a}k},\ and\ \citenamefont {Eschrig}}]{mohn2001correlation}%
  \BibitemOpen
  \bibfield  {author} {\bibinfo {author} {\bibfnamefont {P.}~\bibnamefont {Mohn}}, \bibinfo {author} {\bibfnamefont {C.}~\bibnamefont {Persson}}, \bibinfo {author} {\bibfnamefont {P.}~\bibnamefont {Blaha}}, \bibinfo {author} {\bibfnamefont {K.}~\bibnamefont {Schwarz}}, \bibinfo {author} {\bibfnamefont {P.}~\bibnamefont {Nov{\'a}k}},\ and\ \bibinfo {author} {\bibfnamefont {H.}~\bibnamefont {Eschrig}},\ }\bibfield  {title} {\bibinfo {title} {Correlation induced paramagnetic ground state in \text{FeAl}},\ }\href@noop {} {\bibfield  {journal} {\bibinfo  {journal} {Physical Review Letters}\ }\textbf {\bibinfo {volume} {87}},\ \bibinfo {pages} {196401} (\bibinfo {year} {2001})}\BibitemShut {NoStop}%
\bibitem [{\citenamefont {Oestlin}\ \emph {et~al.}(2017)\citenamefont {Oestlin}, \citenamefont {Vitos},\ and\ \citenamefont {Chioncel}}]{oestlin2017analytic}%
  \BibitemOpen
  \bibfield  {author} {\bibinfo {author} {\bibfnamefont {A.}~\bibnamefont {Oestlin}}, \bibinfo {author} {\bibfnamefont {L.}~\bibnamefont {Vitos}},\ and\ \bibinfo {author} {\bibfnamefont {L.}~\bibnamefont {Chioncel}},\ }\bibfield  {title} {\bibinfo {title} {Analytic continuation-free \text{Green's} function approach to correlated electronic structure calculations},\ }\href@noop {} {\bibfield  {journal} {\bibinfo  {journal} {Physical Review B}\ }\textbf {\bibinfo {volume} {96}},\ \bibinfo {pages} {125156} (\bibinfo {year} {2017})}\BibitemShut {NoStop}%
\bibitem [{\citenamefont {Dastanpour}\ \emph {et~al.}(2025{\natexlab{b}})\citenamefont {Dastanpour}, \citenamefont {Huang}, \citenamefont {Sch{\"o}necker}, \citenamefont {Str{\"o}m}, \citenamefont {Varga}, \citenamefont {Eriksson},\ and\ \citenamefont {Vitos}}]{dastanpour2025magnetocaloric}%
  \BibitemOpen
  \bibfield  {author} {\bibinfo {author} {\bibfnamefont {E.}~\bibnamefont {Dastanpour}}, \bibinfo {author} {\bibfnamefont {S.}~\bibnamefont {Huang}}, \bibinfo {author} {\bibfnamefont {S.}~\bibnamefont {Sch{\"o}necker}}, \bibinfo {author} {\bibfnamefont {V.}~\bibnamefont {Str{\"o}m}}, \bibinfo {author} {\bibfnamefont {L.~K.}\ \bibnamefont {Varga}}, \bibinfo {author} {\bibfnamefont {O.}~\bibnamefont {Eriksson}},\ and\ \bibinfo {author} {\bibfnamefont {L.}~\bibnamefont {Vitos}},\ }\bibfield  {title} {\bibinfo {title} {Magnetocaloric properties of ternary \text{Al}-\text{Mn}-\text{Co} alloys},\ }\href@noop {} {\bibfield  {journal} {\bibinfo  {journal} {Journal of Alloys and Compounds}\ ,\ \bibinfo {pages} {182006}} (\bibinfo {year} {2025}{\natexlab{b}})}\BibitemShut {NoStop}%
\bibitem [{\citenamefont {Stoner}(1938)}]{stoner1938collective}%
  \BibitemOpen
  \bibfield  {author} {\bibinfo {author} {\bibfnamefont {E.~C.}\ \bibnamefont {Stoner}},\ }\bibfield  {title} {\bibinfo {title} {Collective electron ferromagnetism},\ }\href@noop {} {\bibfield  {journal} {\bibinfo  {journal} {Proceedings of the Royal Society of London. Series A. Mathematical and Physical Sciences}\ }\textbf {\bibinfo {volume} {165}},\ \bibinfo {pages} {372} (\bibinfo {year} {1938})}\BibitemShut {NoStop}%
\bibitem [{\citenamefont {Stoner}(1939)}]{stoner1939collective}%
  \BibitemOpen
  \bibfield  {author} {\bibinfo {author} {\bibfnamefont {E.~C.}\ \bibnamefont {Stoner}},\ }\bibfield  {title} {\bibinfo {title} {Collective electron ferromagnetism \text{II}. energy and specific heat},\ }\href@noop {} {\bibfield  {journal} {\bibinfo  {journal} {Proceedings of the Royal Society of London. Series A. Mathematical and Physical Sciences}\ }\textbf {\bibinfo {volume} {169}},\ \bibinfo {pages} {339} (\bibinfo {year} {1939})}\BibitemShut {NoStop}%
\bibitem [{\citenamefont {Ole{\'s}}\ and\ \citenamefont {Stollhoff}(1986)}]{oles1986influence}%
  \BibitemOpen
  \bibfield  {author} {\bibinfo {author} {\bibfnamefont {A.}~\bibnamefont {Ole{\'s}}}\ and\ \bibinfo {author} {\bibfnamefont {G.}~\bibnamefont {Stollhoff}},\ }\bibfield  {title} {\bibinfo {title} {Influence of electron correlation on the stoner parameter and on magnetovolume effect in ferromagnetic transition metals},\ }\href@noop {} {\bibfield  {journal} {\bibinfo  {journal} {Journal of magnetism and magnetic materials}\ }\textbf {\bibinfo {volume} {54}},\ \bibinfo {pages} {1045} (\bibinfo {year} {1986})}\BibitemShut {NoStop}%
\bibitem [{\citenamefont {Fritsche}\ and\ \citenamefont {Weimert}(1998)}]{fritsche1998first}%
  \BibitemOpen
  \bibfield  {author} {\bibinfo {author} {\bibfnamefont {L.}~\bibnamefont {Fritsche}}\ and\ \bibinfo {author} {\bibfnamefont {B.}~\bibnamefont {Weimert}},\ }\bibfield  {title} {\bibinfo {title} {First-principles theory of ferromagnetic and antiferromagnetic order},\ }\href@noop {} {\bibfield  {journal} {\bibinfo  {journal} {physica status solidi (b)}\ }\textbf {\bibinfo {volume} {208}},\ \bibinfo {pages} {287} (\bibinfo {year} {1998})}\BibitemShut {NoStop}%
\bibitem [{\citenamefont {Mackintosh}\ and\ \citenamefont {Andersen}(1980)}]{mackintosh1980chapter}%
  \BibitemOpen
  \bibfield  {author} {\bibinfo {author} {\bibfnamefont {A.}~\bibnamefont {Mackintosh}}\ and\ \bibinfo {author} {\bibfnamefont {O.}~\bibnamefont {Andersen}},\ }\href@noop {} {\bibinfo {title} {Chapter 5.3 in electrons at the fermi surface, edited by m. springford}} (\bibinfo {year} {1980})\BibitemShut {NoStop}%
\bibitem [{\citenamefont {Skriver}(1985)}]{skriver1985crystal}%
  \BibitemOpen
  \bibfield  {author} {\bibinfo {author} {\bibfnamefont {H.~L.}\ \bibnamefont {Skriver}},\ }\bibfield  {title} {\bibinfo {title} {Crystal structure from one-electron theory},\ }\href@noop {} {\bibfield  {journal} {\bibinfo  {journal} {Physical Review B}\ }\textbf {\bibinfo {volume} {31}},\ \bibinfo {pages} {1909} (\bibinfo {year} {1985})}\BibitemShut {NoStop}%
\bibitem [{\citenamefont {Korzhavyi}\ \emph {et~al.}(1995)\citenamefont {Korzhavyi}, \citenamefont {Ruban}, \citenamefont {Abrikosov},\ and\ \citenamefont {Skriver}}]{korzhavyi1995madelung}%
  \BibitemOpen
  \bibfield  {author} {\bibinfo {author} {\bibfnamefont {P.}~\bibnamefont {Korzhavyi}}, \bibinfo {author} {\bibfnamefont {A.}~\bibnamefont {Ruban}}, \bibinfo {author} {\bibfnamefont {I.}~\bibnamefont {Abrikosov}},\ and\ \bibinfo {author} {\bibfnamefont {H.~L.}\ \bibnamefont {Skriver}},\ }\bibfield  {title} {\bibinfo {title} {Madelung energy for random metallic alloys in the coherent potential approximation},\ }\href@noop {} {\bibfield  {journal} {\bibinfo  {journal} {Physical review B}\ }\textbf {\bibinfo {volume} {51}},\ \bibinfo {pages} {5773} (\bibinfo {year} {1995})}\BibitemShut {NoStop}%
\bibitem [{\citenamefont {Born}(1940)}]{born1940stability}%
  \BibitemOpen
  \bibfield  {author} {\bibinfo {author} {\bibfnamefont {M.}~\bibnamefont {Born}},\ }\bibfield  {title} {\bibinfo {title} {On the stability of crystal lattices. \text{I}},\ }in\ \href@noop {} {\emph {\bibinfo {booktitle} {Mathematical Proceedings of the Cambridge Philosophical Society}}},\ Vol.~\bibinfo {volume} {36}\ (\bibinfo {organization} {Cambridge University Press},\ \bibinfo {year} {1940})\ pp.\ \bibinfo {pages} {160--172}\BibitemShut {NoStop}%
\bibitem [{\citenamefont {Pettifor}(1992)}]{pettifor1992theoretical}%
  \BibitemOpen
  \bibfield  {author} {\bibinfo {author} {\bibfnamefont {D.}~\bibnamefont {Pettifor}},\ }\bibfield  {title} {\bibinfo {title} {Theoretical predictions of structure and related properties of intermetallics},\ }\href@noop {} {\bibfield  {journal} {\bibinfo  {journal} {Materials science and technology}\ }\textbf {\bibinfo {volume} {8}},\ \bibinfo {pages} {345} (\bibinfo {year} {1992})}\BibitemShut {NoStop}%
\bibitem [{\citenamefont {Zhang}\ \emph {et~al.}(2018)\citenamefont {Zhang}, \citenamefont {Sun}, \citenamefont {Lu}, \citenamefont {Dong}, \citenamefont {Ding}, \citenamefont {Wang},\ and\ \citenamefont {Vitos}}]{zhang2018elastic}%
  \BibitemOpen
  \bibfield  {author} {\bibinfo {author} {\bibfnamefont {H.}~\bibnamefont {Zhang}}, \bibinfo {author} {\bibfnamefont {X.}~\bibnamefont {Sun}}, \bibinfo {author} {\bibfnamefont {S.}~\bibnamefont {Lu}}, \bibinfo {author} {\bibfnamefont {Z.}~\bibnamefont {Dong}}, \bibinfo {author} {\bibfnamefont {X.}~\bibnamefont {Ding}}, \bibinfo {author} {\bibfnamefont {Y.}~\bibnamefont {Wang}},\ and\ \bibinfo {author} {\bibfnamefont {L.}~\bibnamefont {Vitos}},\ }\bibfield  {title} {\bibinfo {title} {Elastic properties of {$\mathrm{Al_{x}CrMnFeCoNi (0 \leq x \leq 5}$} high-entropy alloys from $ab$ $initio$ theory},\ }\href@noop {} {\bibfield  {journal} {\bibinfo  {journal} {Acta Materialia}\ }\textbf {\bibinfo {volume} {155}},\ \bibinfo {pages} {12} (\bibinfo {year} {2018})}\BibitemShut {NoStop}%
\bibitem [{\citenamefont {Temesi}\ \emph {et~al.}(2024)\citenamefont {Temesi}, \citenamefont {Varga}, \citenamefont {Chinh},\ and\ \citenamefont {Vitos}}]{temesi2024ductility}%
  \BibitemOpen
  \bibfield  {author} {\bibinfo {author} {\bibfnamefont {O.~K.}\ \bibnamefont {Temesi}}, \bibinfo {author} {\bibfnamefont {L.~K.}\ \bibnamefont {Varga}}, \bibinfo {author} {\bibfnamefont {N.~Q.}\ \bibnamefont {Chinh}},\ and\ \bibinfo {author} {\bibfnamefont {L.}~\bibnamefont {Vitos}},\ }\bibfield  {title} {\bibinfo {title} {Ductility index for refractory high entropy alloys},\ }\href@noop {} {\bibfield  {journal} {\bibinfo  {journal} {Crystals}\ }\textbf {\bibinfo {volume} {14}},\ \bibinfo {pages} {838} (\bibinfo {year} {2024})}\BibitemShut {NoStop}%
\bibitem [{\citenamefont {Chen}\ \emph {et~al.}(2016)\citenamefont {Chen}, \citenamefont {Yao}, \citenamefont {Zhang}, \citenamefont {Luo}, \citenamefont {Zhang},\ and\ \citenamefont {Han}}]{chen2016elastic}%
  \BibitemOpen
  \bibfield  {author} {\bibinfo {author} {\bibfnamefont {Y.}~\bibnamefont {Chen}}, \bibinfo {author} {\bibfnamefont {Z.}~\bibnamefont {Yao}}, \bibinfo {author} {\bibfnamefont {P.}~\bibnamefont {Zhang}}, \bibinfo {author} {\bibfnamefont {X.}~\bibnamefont {Luo}}, \bibinfo {author} {\bibfnamefont {Z.}~\bibnamefont {Zhang}},\ and\ \bibinfo {author} {\bibfnamefont {P.}~\bibnamefont {Han}},\ }\bibfield  {title} {\bibinfo {title} {Elastic constants and properties of \text{B2-type} \text{FeAl} and \text{Fe-Cr-Al} alloys from first-principles calculations},\ }in\ \href@noop {} {\emph {\bibinfo {booktitle} {2nd Annual International Conference on Advanced Material Engineering (AME 2016)}}}\ (\bibinfo {organization} {Atlantis Press},\ \bibinfo {year} {2016})\ pp.\ \bibinfo {pages} {380--386}\BibitemShut {NoStop}%
\bibitem [{\citenamefont {Pagare}\ \emph {et~al.}(2015)\citenamefont {Pagare}, \citenamefont {Jain}, \citenamefont {Abraham},\ and\ \citenamefont {Sanyal}}]{pagare2015first}%
  \BibitemOpen
  \bibfield  {author} {\bibinfo {author} {\bibfnamefont {G.}~\bibnamefont {Pagare}}, \bibinfo {author} {\bibfnamefont {E.}~\bibnamefont {Jain}}, \bibinfo {author} {\bibfnamefont {J.~A.}\ \bibnamefont {Abraham}},\ and\ \bibinfo {author} {\bibfnamefont {S.}~\bibnamefont {Sanyal}},\ }\bibfield  {title} {\bibinfo {title} {First-principles theoretical prediction of structural and elastic properties of some \text{AlX} intermetallics},\ }in\ \href@noop {} {\emph {\bibinfo {booktitle} {AIP Conference Proceedings}}},\ Vol.\ \bibinfo {volume} {1675}\ (\bibinfo {organization} {AIP Publishing LLC},\ \bibinfo {year} {2015})\ p.\ \bibinfo {pages} {030058}\BibitemShut {NoStop}%
\bibitem [{\citenamefont {Ponomareva}\ \emph {et~al.}(2014)\citenamefont {Ponomareva}, \citenamefont {Vekilov},\ and\ \citenamefont {Abrikosov}}]{ponomareva2014effect}%
  \BibitemOpen
  \bibfield  {author} {\bibinfo {author} {\bibfnamefont {A.}~\bibnamefont {Ponomareva}}, \bibinfo {author} {\bibfnamefont {Y.~K.}\ \bibnamefont {Vekilov}},\ and\ \bibinfo {author} {\bibfnamefont {I.}~\bibnamefont {Abrikosov}},\ }\bibfield  {title} {\bibinfo {title} {Effect of \text{Re} content on elastic properties of \text{B2 NiAl} from $ab$ $initio$ calculations},\ }\href@noop {} {\bibfield  {journal} {\bibinfo  {journal} {Journal of alloys and compounds}\ }\textbf {\bibinfo {volume} {586}},\ \bibinfo {pages} {S274} (\bibinfo {year} {2014})}\BibitemShut {NoStop}%
\bibitem [{\citenamefont {Huang}\ \emph {et~al.}(2018{\natexlab{b}})\citenamefont {Huang}, \citenamefont {Li}, \citenamefont {Holmstr{\"o}m},\ and\ \citenamefont {Vitos}}]{huang2018strengthening}%
  \BibitemOpen
  \bibfield  {author} {\bibinfo {author} {\bibfnamefont {S.}~\bibnamefont {Huang}}, \bibinfo {author} {\bibfnamefont {W.}~\bibnamefont {Li}}, \bibinfo {author} {\bibfnamefont {E.}~\bibnamefont {Holmstr{\"o}m}},\ and\ \bibinfo {author} {\bibfnamefont {L.}~\bibnamefont {Vitos}},\ }\bibfield  {title} {\bibinfo {title} {Strengthening induced by magnetochemical transition in \text{Al-doped Fe-Cr-Co-Ni} high-entropy alloys},\ }\href@noop {} {\bibfield  {journal} {\bibinfo  {journal} {Physical Review Applied}\ }\textbf {\bibinfo {volume} {10}},\ \bibinfo {pages} {064033} (\bibinfo {year} {2018}{\natexlab{b}})}\BibitemShut {NoStop}%
\end{thebibliography}%

\end{document}